\documentclass[11pt]{article}
\usepackage[marginratio=1:1,height=600pt,width=460pt,tmargin=100pt]{geometry}

\usepackage{graphicx}
\usepackage[utf8]{inputenc} 
\usepackage[T1]{fontenc}    
\usepackage[hidelinks, colorlinks=true,citecolor=blue]{hyperref} 
\usepackage{url}
\usepackage{booktabs}
\usepackage{amsfonts}
\usepackage{nicefrac}
\usepackage{microtype}
\usepackage{xcolor}
\usepackage[normalem]{ulem}
\usepackage{authblk}
\usepackage{amsthm,amsmath,amsfonts,amssymb}
\usepackage{mathtools}
\usepackage{mathrsfs}
\usepackage{color}
\usepackage{natbib}
\usepackage{caption}
\usepackage{subcaption}
\usepackage{bm}
\usepackage[flushleft]{threeparttable}
\usepackage[shortlabels]{enumitem}
\usepackage{wrapfig}
\setlist{leftmargin=6mm}
\usepackage{multirow}
\usepackage{titlesec}
\usepackage{setspace}
\usepackage{pifont} % Dingbats for checkmarks and crosses
\usepackage{algorithm}
\usepackage{algorithmic}
\newlength\myindent
\newtheorem{conjecture*}{Conjecture}
\theoremstyle{remark}

\allowdisplaybreaks

\title{Spatio-Temporal-Network Point Processes for Modeling \\ Crime Events with Landmarks}

\author[1]{Zheng Dong}
\author[2]{Jorge Mateu}
\author[1]{Yao Xie\footnote{Email: yao.xie@isye.gatech.edu}}
\affil[1]{{\small H. Milton Stewart School of Industrial and Systems Engineering, Georgia Institute of Technology}\vspace{2pt}}
\affil[2]{{\small Department of Mathematics, Universitat Jaume I, Castell\'o de la Plana, Valencia, Spain}\vspace{-0.3in}}
\date{}

\begin{document}

\maketitle

\begin{abstract}

Self-exciting point processes are widely used to model the contagious effects of crime events living within continuous geographic space, using their occurrence time and locations.
However, in urban environments, most events are naturally constrained within the city's street network structure, and the contagious effects of crime are governed by such a network geography. Meanwhile, the complex distribution of urban infrastructures also plays an important role in shaping crime patterns across space. We introduce a novel spatio-temporal-network point process framework for crime modeling that integrates these urban environmental characteristics by incorporating self-attention graph neural networks. Our framework incorporates the street network structure as the underlying event space, where crime events can occur at random locations on the network edges. To realistically capture criminal movement patterns, distances between events are measured using street network distances. We then propose a new mark for a crime event by concatenating the event's crime category with the type of its nearby landmark, aiming to capture how the urban design influences the mixing structures of various crime types. A graph attention network architecture is adopted to learn the existence of mark-to-mark interactions.
Extensive experiments on crime data from Valencia, Spain, demonstrate the effectiveness of our framework in understanding the crime landscape and forecasting crime risks across regions.
\end{abstract}

%%%%%%%%%%%%%%%%%%%%%%%%%%%%%%%%%%%%%%%%%%%%%%
%% Please use \tableofcontents for articles %%
%% with 50 pages and more                   %%
%%%%%%%%%%%%%%%%%%%%%%%%%%%%%%%%%%%%%%%%%%%%%%
%\tableofcontents

%%%%%%%%%%%%%%%%%%%%%%%%%%%%%%%%%%%%%%%%%%%%%%
%%%% Main text entry area:

\section{Introduction}

Self-exciting point processes \citep{reinhart2018review} have been used in crime modeling with several successful attempts on burglary \citep{mohler2011self}, gang violence \citep{zipkin2014cops}, gunshot incidents \citep{dong2024atlanta}, and terrorism data \citep{porter2012self}. The statistical structure of a self-exciting process is well-suited to characterize both the endogenous crime rates and the contagious pattern observed in crime data \citep{johnson2008repeat, mohler2013modeling, loeffler2018gun}. Specifically, it models the intensity of crime events using a background event rate and a so-called \textit{influence kernel} that plays a pivotal role in capturing the contagious effect of an observed crime event on future crime events in nearby neighborhoods.

The dynamics of crime contagion are particularly complex within urban settings, influenced heavily by the geographic layout of the city. While crimes occur in a continuous space (\textit{e.g.}, within a city area measured by longitude and latitude), they are mostly confined to the street networks, influencing both the escape routes of criminals and the spatial distribution of crime \citep{rossmo1999geographic}.
Early research \citep{mohler2011self} also suggests that crime's contagious effects propagate along these street networks instead of dispersing freely, as evidenced by a fitted non-parametric influence kernel from real crime events.
In this situation, traditional point processes with Euclidean distance-based influence kernels \citep{mohler2014marked, reinhart2018self, zhuang2019semiparametric} often fall short, necessitating an adjusted influence kernel that respects this urban constraint.

Another factor contributing to the complexity of urban crime dynamics is the diversity of the surrounding urban environments where different crimes occur. Diverse land uses, ranging from commercial to residential areas, influence the types and prevalence of criminal activities, each fostering unique interactions between potential offenders and victims \citep{fleming1994exploring, stucky2009land, kinney2008crime}.
For instance, commercial areas can host a variety of legitimate (shopping, working, eating, etc.) and criminal (shoplifting, picking pockets, etc.) activities during business hours, creating specific crime patterns that are very different from those in other regions \citep{kinney2008crime}.
While previous studies have shown the effectiveness of fine-crafted point process models in understanding the landscapes of various crime types across different regions \citep{mohler2014marked, linderman2014discovering}, there remains a gap in these models' capability to integrate the information of urban land uses, limiting their explanatory power regarding the relationship between specific urban surroundings and crime patterns.
% requiring advanced models capable of integrating contextual information such as the geographic conditions and crime types.
% For example, the daily movements of vehicles – their concentration in work venues, shopping areas, or residential neighborhoods at different times of the day – provide a dynamic distribution of targets for vehicle theft \citep{fleming1994exploring}.

In this paper, we introduce a novel spatio-temporal-network point process model tailored for analyzing crime within urban street networks. This model uniquely incorporates the structure of city street networks and adopts a street-network-based distance metric that aligns more closely with the actual movement patterns of criminals compared to traditional Euclidean metrics, providing a realistic depiction of crime patterns within a networked urban environment.
To integrate the contextual data of urban land uses into the model, we craft a special mark for each crime event, which considers the information about nearby landmarks such as banks, restaurants, and supermarkets. Using the concept of urban functional zones \citep{yuan2014discovering} that segment the entire city area based on the landmarks, we extend the traditional mark of a crime event, typically the category of the crime \citep{mohler2014marked, reinhart2018self} (\textit{e.g.}, burglary, larceny, robbery, etc.), into a new mark that contains both the crime and landmark categories. Such an event mark allows for direct analysis of the impact of specific urban surroundings on local crime patterns.

% Specifically, we first collect the locations of different types of landmarks in the city, such as banks, restaurants, and supermarkets. Based on these landmarks, we can segment the entire city area into different urban functional zones \citep{yuan2014discovering}. We then extend the traditional mark of a crime event, typically the category of the crime \citep{mohler2014marked, reinhart2018self} (\textit{e.g.}, burglary, larceny, robbery, etc.), by integrating the type of the urban functional zone the event falls in. Such an event mark allows for a straightforward analysis of the impact of specific urban surroundings on crime patterns.

The design of our influence kernel jointly considers the time, location, and mark information of crime events. A temporal kernel and a street distance-based spatial kernel characterize how previous crimes influence future ones over time and space, respectively. 
Moreover, we introduce a novel mark network captured through graph neural networks (GNNs), which assesses interactions between different crime events based on their marks. This GNN framework predicts potential linkages between different marks while considering their intrinsic similarities. By capturing these intricate relationships, our model facilitates a deeper understanding of crime clustering and propagation. 
Tested extensively with real crime data from Valencia, Spain, our model has proven highly effective in capturing the dynamic landscape of urban crime and predicting crime risk across the city, offering significant improvements over existing methodologies.

% to capture the interactions between crime events with different marks. 
% We leverage graph neural networks (GNNs) within our influence kernel to model the complex interactions among different event marks. The GNN framework enables us to estimate the topology of the mark network (the existence of interactions between marks) while considering the intrinsic similarities of the event marks. The capture of the mark interactions helps identify groups of closely related crime events in downstream tasks. Our comprehensive approach is validated through extensive numerical experiments using real crime data from Valencia, Spain. The results demonstrate that our model not only effectively captures the dynamic landscape of urban crime but also significantly enhances the accuracy of crime risk predictions across the city.

% The learning of the mark interaction network is achieved by incorporating graph neural networks (GNNs) in the influence kernel. The GNNs are commonly used to learn the network topology between nodes by considering the nodal features, which we adopt here to estimate the existence of interactions between different crime categories and landmarks through their feature vectors. The interaction strength is learned separately. The incorporation of GNN and mark features plays an important role in our influence kernel because crime events with distinct marks can share intrinsic similarities if they have the same crime or landmark category. 

The paper is organized as follows. The rest of this section reviews related literature. Section~\ref{sec:data} introduces the crime and landmark data sets collected in Valencia that motivate our model. Section~\ref{sec:data-processing} presents the data-processing techniques that define the format of the discrete event data with marks. Section~\ref{sec:point-process-modeling} introduces our spatio-temporal-network point-process model with graph neural networks, which is learned using the estimation strategy in Section~\ref{sec:model-estimation}. Finally, in Section~\ref{sec:results}, we present the results using our model on the real crime data in Valencia and a comparison with baselines. The paper ends with some further discussion.

% Within this approach, a classical possibility is to interpret the interdependencies between time series (encapsulated, for instance, in cross-correlation matrices) as the weighted edges of a graph whose nodes label each time series, yielding so-called functional networks that have been used fruitfully and extensively in different fields such as neuroscience or finance.

% By jointly consider the multiple level of information, we allow for a more interpretable and informative characterization of the interactions between different crime incidents.

% novel use of graph attention neural networks together with point process kernel, to achieve learning of topology (finding the linkages between different categories and landmarks); which can help to identify subgroups

\subsection{Related work}

% \textcolor{red}{Add references on point processes; point processes on graph, from both statistics, and machine learning}

Our research is placed within the domain of predictive policing \citep{perry2013predictive}, which includes four general categories: methods for predicting crimes \citep{chainey2008utility, neill2007detecting, wang2012spatio}, methods for predicting offenders \citep{bonta1998prediction, grann1999psychopathy}, methods for predicting perpetrators' identities \citep{lev2004posttraumatic, tarzia2018exploring}, and methods for predicting victims of crime \citep{gottfredson1981etiology, russo2013criminal}. Our study belongs to the first category, which aims to forecast places and times with an increased crime risk. Unlike the other three categories that require the collection of extensive information about crime incidents, such as police reports, to identify certain individuals or groups that may get involved in criminal activities, the prediction of spatio-temporal occurrences of crime can be mainly achieved by leveraging historical crime data without the necessary access to sensitive information.

Many mathematical models have been used to understand the complex phenomenon of crime; a family of those includes tools that aim to detect potential hotspots based on empirical observations of spatial clusters of crime incidents \citep{levine2002spatial, bowers2004prospective, chainey2008utility}.
However, most hotspot modeling approaches do not consider the temporal dynamics of the hotspot, despite some exploring the overall evolution of hotspots rather than focusing on individual events \citep{short2008statistical}.
Other models use regression-based methods \citep{meera1995determinants, kennedy2011risk, kennedy2016vulnerability} to quantitatively assess the effects of different factors on the total number of crimes in a specific region.
Along this line, recent works \citep{hessellund2022second, hessellund2022semiparametric, xu2023semiparametric} have developed  semiparametric frameworks for fitting and testing spatial covariate effects on the spatial intensity of crime events.
Interpretable results on covariate effects from regression models can potentially help with more targeted interventions. These methods usually require the collection of contextual information, such as demographic and socioeconomic data, to establish the regression models.
In contrast, our method models discrete crime event data to capture the spatio-temporal near-repeat effect of crime, and enables fine-grained prediction and risk evaluation over the street network in a data-driven manner.

In recent decades, there has been a substantial body of research \citep{kinney2008crime, johnson2010permeability, groff2011exploring, weisburd2012criminology, xu2017shooting} examining the relationship between urban land use and crime patterns. These studies have pinpointed environmental characteristics linked to increased crime risks in specific urban settings. 
Our approach differs from the aggregated-statistics-based analysis often taken in such research \citep{fleming1994exploring, kinney2008crime, stucky2009land, browning2010commercial, xu2017shooting}. Instead, we model the spatio-temporal crime patterns through the lens of individual crime incidents, providing a dynamic perspective for explaining the crime and integrating effective surveillance.

The application of self-exciting point processes, motivated by the modeling of earthquake occurrences in seismology \citep{ogata1988statistical}, has been widely explored to characterize the dynamics of criminal activities \citep{mohler2011self, lewis2012self, mohler2013modeling, reinhart2018review, zhuang2019semiparametric, zhu2022spatiotemporal}. Previous attempts demonstrate the effectiveness of point processes in modeling crime using residential burglary data in Los Angeles \citep{mohler2011self}, civilian death reports in Iraq \citep{lewis2012self}, and gunshot data in Washington, DC \citep{loeffler2018gun}. 
% while they only consider the temporal patterns of the crime incidents. 
Later approaches \citep{mohler2014marked, reinhart2018self, zhu2022spatiotemporal} improve point process models for crime by incorporating the events' type, location, and textual information to capture complex crime patterns in different modeling tasks. 
Compared with them, our approach extends the traditional modeling of crime events in Euclidean space by adopting a network distance between crime events that is more realistic to estimate the travel distance of criminals in the urban environment. A recent paper considers crime events on linear street network \citep{d2024self} that focus on improving the estimation of the non-parametric influence kernel and event intensity. Our model differs from theirs by considering an influence kernel that can leverage multiple levels of information.

% not only influence the crime rates but also create different opportunity structures for crime incidents by shaping the mixture of motivated offenders and potential victims \citep{stucky2009land}. 
% For example, the daily movements of vehicles – their concentration in work venues, shopping areas, or residential neighborhoods at different times of the day – provide a dynamic distribution of targets for vehicle theft \citep{fleming1994exploring}. 
% Therefore, it is of great scientific and practical meaning to investigate certain types of crime while considering their relationship with particular urban functional zones.

% Depending on the context of the local land uses and urban structures, a specific urban functional region would also have a specific crime pattern that is very different from those in other regions \citep{kinney2008crime}. At a grander scale, a shopping centre or shopping district has busy hours when a broad range of people are present for many reasons – shopping, working, meeting friends, eating, drinking or walking through. During such busy times the crowd can also include beggars, buskers, and thieves who are there for shoplifting, for picking pockets, for credit card fraud or for stealing from motor vehicles. Shopping centers can support a variety of legitimate and criminal activities during business hours.

% \paragraph{kernel modeling in self-exciting pp for crime}
% \citep{mohler2011self}

A number of studies have considered the problem of modeling discrete events observed within network structures using self-exciting point processes in addition to modeling crime incidents. Most of them \citep{liao2022tides, fang2023group, cai2024latent, sanna2024graph} only model the temporal occurrences of the events that come from the nodes in the networks. 
The work of network Hawkes \citep{linderman2014discovering} adopts a similar decomposed representation as ours to capture the event mark interactions. However, their approach includes the estimation of a binary random matrix using Gibbs sampling, which is fundamentally different and more complicated, and the events' location information is processed on an aggregated level. 
Other studies have extended to multilayer network settings \citep{kivela2014multilayer, li2023stochastic, liu2025dynamic, liu2025network}, where nodes are connected through multiple types of network relationships. For example, \citet{cho2013latent} addresses the missing data problem in spatio-temporal social networks with geographically distributed nodes connected via a social network.
While they restrict events to nodes and emphasize node-level relationships, our framework models events along edges and directly captures event dependencies over multiple network topologies.
% The missing data problem also exists in criminology, such as detecting unreported crimes, which is not the focus of our study. 

% combines the Hawkes processes and random network models to infer implicit network structure from observed events through a similar decomposed representation for the mark interactions as ours. However, their approach includes the estimation of a binary random matrix using Gibbs sampling which is fundamentally different and more complicated, and the events' location information is processed on an aggregated level. 

% Network distance between two events at random locations on the network \citep{wei2020distance}.

% \paragraph{Graph neural networks}

Last but not least, incorporating neural networks in point process models has recently been a popular research topic \citep{shchur2021neural}. Various neural point processes focus on leveraging recurrent neural networks (RNNs) \citep{du2016recurrent,mei2017neural} or Transformer structure \citep{zuo2020transformer, zhang2020self} to encode the historical information. Compared with our method, they did not consider the statistical framework of the self-exciting point process and often lacked model interpretability. Another line of work \citep{cheng2025deep, dong2023spatiotemporal, dong2023non, zhu2021imitation, zhu2021neural, zhu2021deep} focuses on representing the influence kernel using neural networks, allowing for the modeling of a wider range of complex event dynamics such as non-stationary and inhibiting effects. However, they do not consider contextual information such as the latent network structure or mark features. Graph neural networks have been extensively developed within the machine learning community. Nevertheless, their application in point processes has received scant investigation. Two recent works use message-passing GNNs in point processes \citep{xia2022graph, wu2020modeling} for the task of temporal link prediction rather than modeling discrete marked events.
Another concurrent study of graph point processes \citep{dong2023deep} shares similarities with ours by approximating influence kernels using graph neural networks. However, they do not consider the spatial aspect of the data.

\section{Data description}
\label{sec:data}

\begin{figure}[!t]
%\vspace{-.15in}
\centering
\begin{subfigure}[h]{.32\linewidth}
\includegraphics[width=\linewidth]{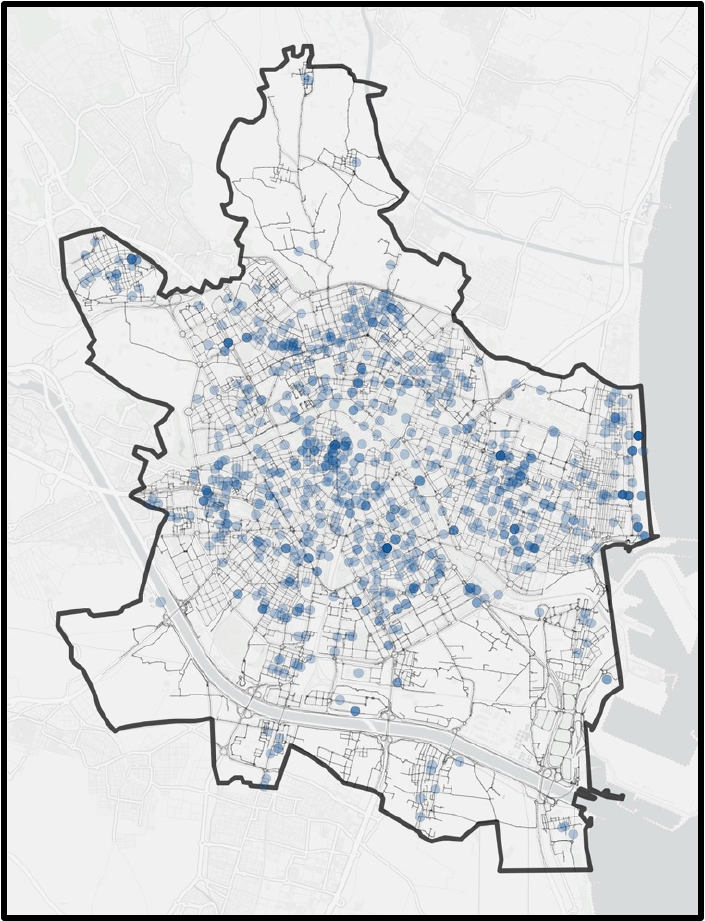}
\caption{Crime, July 2016}
\end{subfigure}
\begin{subfigure}[h]{.32\linewidth}
\includegraphics[width=\linewidth]{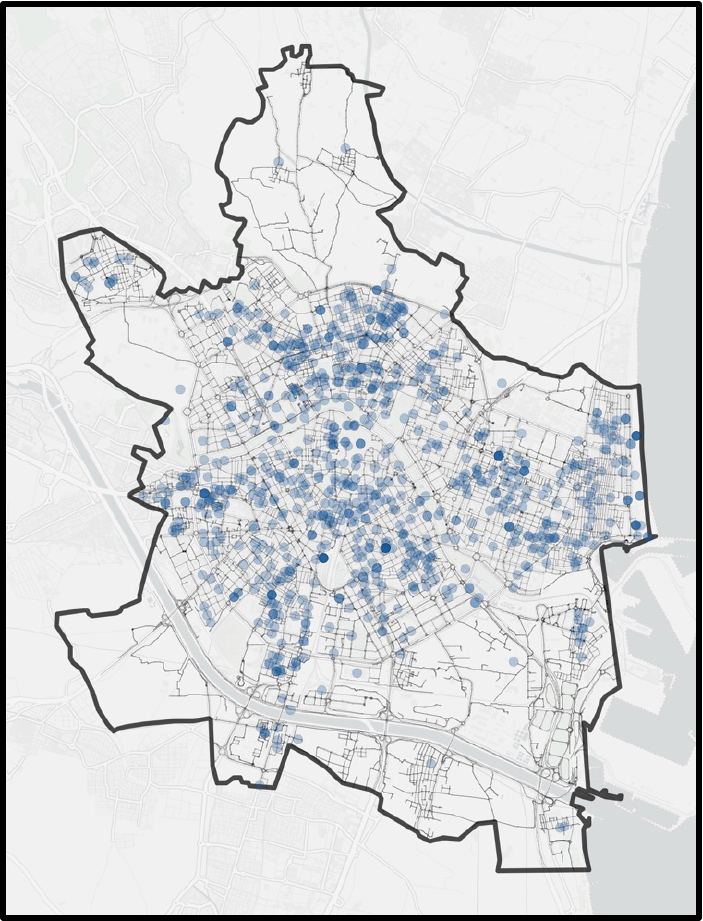}
\caption{Crime, October 2018}
\end{subfigure}
\begin{subfigure}[h]{.32\linewidth}
\includegraphics[width=\linewidth]{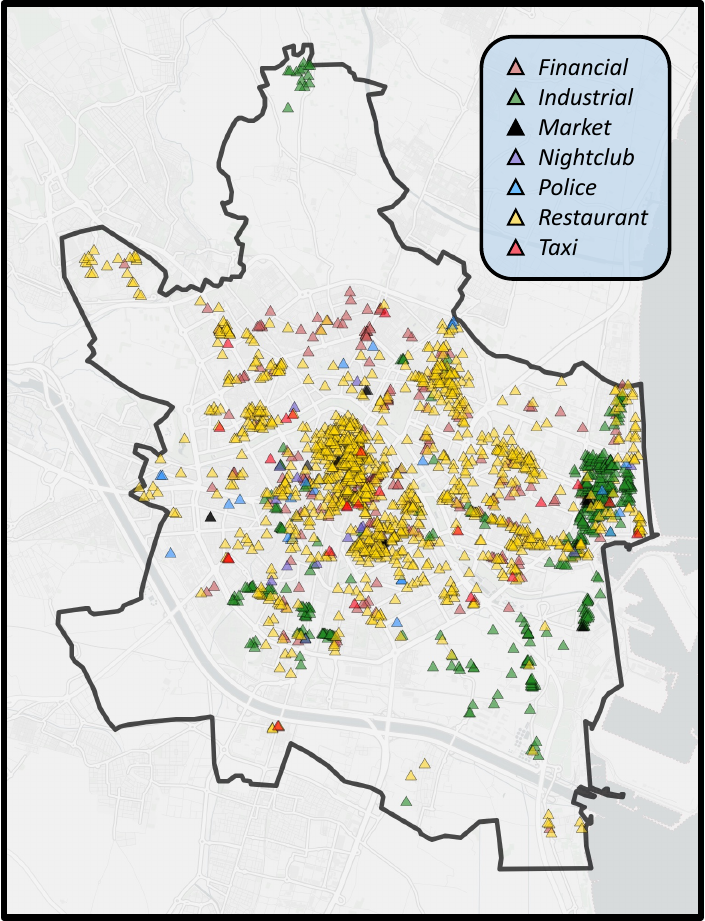}
\caption{City landmarks}
\end{subfigure}
\vspace{-0.1in}
\caption{Two snapshots of crime incidents that happened on the street network in the city of Valencia (Spain) at different times are shown in (a) and (b). The blue dots represent the recorded incidents by the local police department, with a deeper color indicating multiple incidents within a small area. The grey lines represent the street network in the city of Valencia. (c) Locations of city landmarks. Each triangle represents one landmark, with different colors suggesting different landmark types.
}
\label{fig:data-visualization-landmark}
\vspace{-0.1in}
\end{figure}

% \begin{figure}[!t]
% %\vspace{-.15in}
% \centering
% \begin{subfigure}[h]{.32\linewidth}
% \includegraphics[width=\linewidth]{Agresion.pdf}
% \caption{{\it Agresi\'{o}n} (assault)}
% \end{subfigure}
% \begin{subfigure}[h]{.32\linewidth}
% \includegraphics[width=\linewidth]{Sustraccion.pdf}
% \caption{{\it Sustracci\'{o}n} (theft)}
% \end{subfigure}
% \begin{subfigure}[h]{.32\linewidth}
% \includegraphics[width=\linewidth]{Otros.pdf}
% \caption{{\it Otros} (others)}
% \end{subfigure}
% \caption{Distributions of three types of crime incidents over the street network. The distributions are evaluated using all the incidents from 2015 to 2019. The red color depth indicates the incident density across different streets.
% }
% \label{fig:crime-spatial-distribution}
% % \vspace{-0.1in}
% \end{figure}

% The time range is from 2010 to 2020, so 10 complete years and we have a total of 90247 events, splitted in "Agresion" (55610 cases), "Sustraccion" (25342 cases), "AlarmasMujer" (454 cases) and "Otros" (8841 cases). These 4 types have to do with different types of thefts or robberies in the streets: "Agresion" means a theft after hitting a person, "Sustraccion" means a smooth theft with no force used, "AlarmasMujer" has to do with a theft to a woman with violence, "Otros" means other thefts or robberies that can not be considered as in the previous categories.

The crime data in this study is collected by the local police department in Valencia (Spain), a town located along the Mediterranean coast with more than 1.5 million inhabitants. The data set records thefts and robberies over five years from 2015 to 2019, including a total of $47,125$ crime events. Each record contains comprehensive information about one event, including time, location (measured in longitude and latitude), and the crime category. 
The recorded events are categorized into three distinct types, including: (i) \textit{Assault} (\textit{Agresi\'{o}n}, in its source name) referring to thefts involving physical assault, (ii) \textit{Subtraction (Sustracci\'{o}n)} referring to thefts executed smoothly without the use of force, and (iii) \textit{Others (Otros)} referring to other types of street thefts or robberies not included in the previous categories.

The data set uniquely focuses on crimes that occurred on city streets, as emphasized by the local police department. To support our data analysis, we acquire street network data within the Valencia city boundary from the OpenStreetMap database \citep{OpenStreetMap}.
Fig~\ref{fig:data-visualization-landmark}(a) and (b) provide visual snapshots of the recorded crime events scattered across Valencia's street network in July 2016 and October 2018, respectively.
% ``AlarmasMujer'' has to do with theft to a woman with violence

Additionally, to investigate the relationship between the patterns of the reported crimes and the surrounding urban environment, we collect the location information of $1,975$ city landmarks in Valencia, categorized into seven types: financial, industrial, market, nightclub, police, restaurant, and taxi. Fig~\ref{fig:data-visualization-landmark}(c) visualizes the spatial distribution of these landmarks, with different colors indicating different categories. 
The landmark data were obtained from the last official release prior to 2015 by the Valencia city government, originally compiled using the Google Maps API. This dataset remained unchanged during our study period (2015–2019), as no official updates were released in those years. The next comprehensive update occurred in 2021, after our study window. Thus, using this dataset ensures temporal consistency across the five-year analysis period.

% \begin{figure}[!tb]
% %\vspace{-.15in}
% \centering

% \begin{subfigure}[h]{.6\linewidth}
% \includegraphics[width=\linewidth]{crime-ldmk_type_data_vis.pdf}
% % \caption{Vanilla Hawkes}
% \end{subfigure}

% \caption{Case distribution with different crime-landmark labels. The color depth indicates the case density
% }
% \label{fig:crime-ldmk-type-vis}
% % \vspace{-0.1in}
% \end{figure}

\section{Data processing}
\label{sec:data-processing}

We first present the processing strategies for our crime data set, as they play an important role in characterizing the latent and complex correlation structure presented in the events.
Consider a sequence of $n$ reported crime events in Valencia. Denoting each event as a tuple, the entire sequence of events can be represented as
\begin{equation}
    (t_1, s_1, c_1), (t_2, s_2, c_2), \dots,(t_n, s_n, c_n).
    \label{eq:raw-data}
\end{equation}
For the $i$-th event, $t_i \in [0, T]$ represents the time of incident occurrence, $s_i \in \mathcal{S} \subseteq \mathbb{R}^2$ denotes the location of the incident, measured in longitude and latitude coordinates, where $\mathcal{S}$ denotes the geographical area covered by the city of Valencia, and 
$c_i \in \mathscr{C} \coloneqq \{1, 2, 3\}$ denotes the crime category of the $i$-th incident, with $1, 2, \text{and}~ 3$ representing \textit{Assault}, \textit{Subtraction}, and \textit{Others}, respectively. All the events are temporally ordered, \textit{i.e.}, $0 \leq t_1 < t_2 < \dots < t_{n} \leq T$. 
% The entire sequence can be denoted as $\mathcal{H}_T = \{(t_i, s_i, c_i)\}_{i=1}^{n}$.
We also introduce the notation for seven landmark categories as $\mathscr{L} \coloneqq [1:7]$. Values from $1$ to $7$ correspond to the landmark categories of financial, industrial, market, nightclub, police, restaurant, and taxi, respectively.

\subsection{Urban functional zone identification}

% It is worth noting that the landmark locations and density are highly correlated with the crime rates over space. 
% The distribution of urban land uses and the locations of different kinds of city landmarks are critically important in criminological studies. 
Urban areas with different facilities and functionalities, known as \textit{urban functional zones} \citep{yuan2014discovering}, can have different crime patterns based on the citizen activities exhibited in those areas \citep{kinney2008crime}. 
For instance, commercial or public places attract more human activities and, potentially, more crime and disorder events \citep{andresen2007location, 10.1093/oxfordhb/9780190279707.013.14}.
The identification of the urban functional zones is critical for implementing targeted crime prevention strategies and mitigating potential hotspots.

% based on the collected landmark data, according to which we can investigate the varying likelihood of crime events in different environmental characteristics.

\begin{figure}[!t]
%\vspace{-.15in}
\centering
\begin{subfigure}[h]{.85\linewidth}
\includegraphics[width=\linewidth]{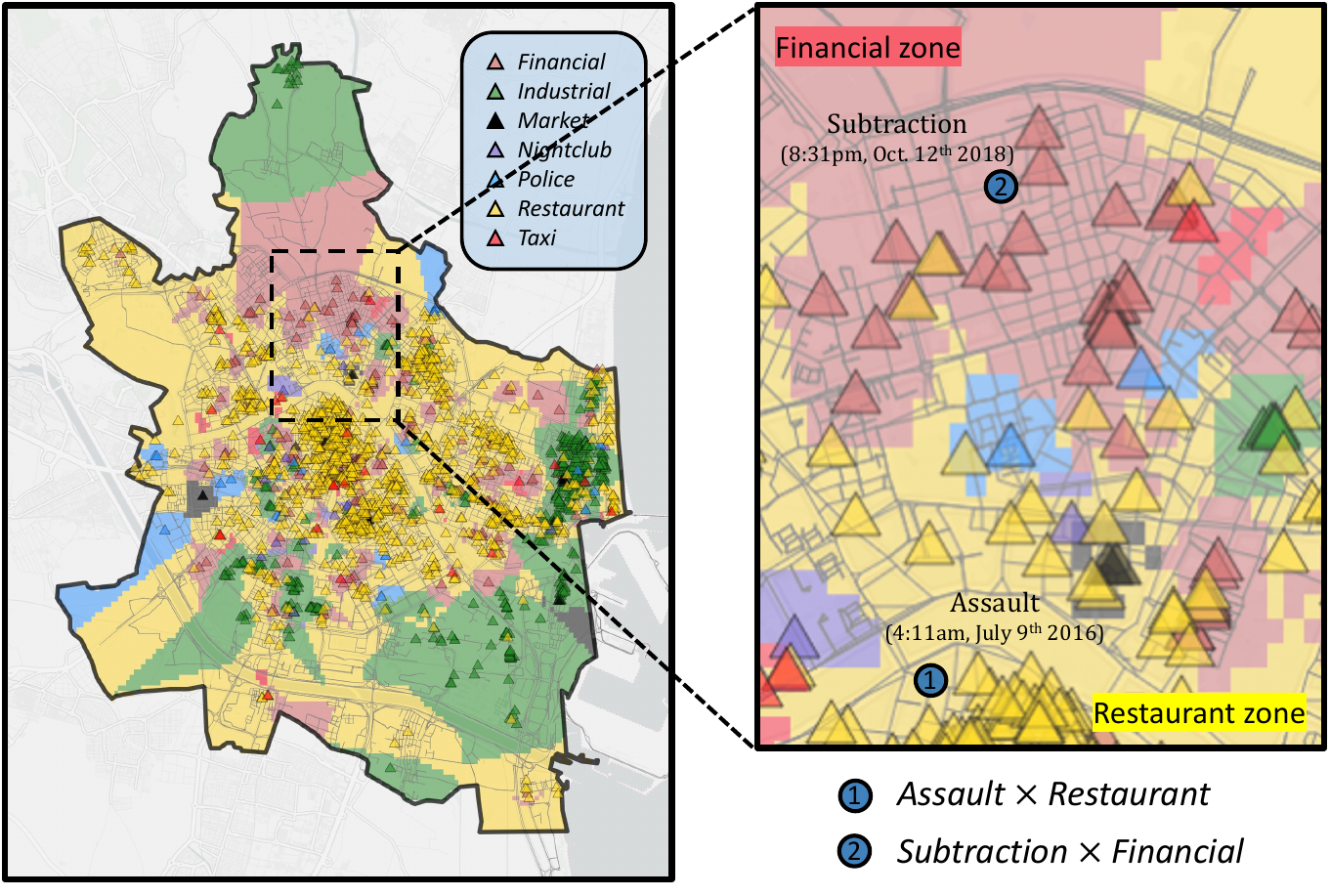}
\end{subfigure}
\caption{Partition of Valencia city area into various urban functional zones, and labeling of crime incidents with the joint categories of crime and landmark. \textit{Left}: the entire city area is divided into zones with different functionalities based on the spatial proximity to different city landmarks. Each zone is highlighted in the same color as the corresponding landmark category. \textit{Right}: The labeling of each crime incident is jointly determined by its crime category and the functional zone it falls in.
}
\label{fig:join-crime-landmark}
\vspace{-0.1in}
\end{figure}

In our study, we partition the entire city area of Valencia into various urban functional zones based on the 1,975 city landmarks. This approach aligns with the point-of-interest (POI) method commonly referenced in the literature \citep{gao2017extracting, hu2019identification, long2015discovering, yuan2014discovering}, which involves geographic entities that can be abstracted as points for zone identification, such as schools, banks, companies, restaurants, and supermarkets \citep{jiang2015mining}. 
Specifically, we use the $k$-nearest neighbors algorithm to identify different functional zones based on their proximity to the city landmarks.
For a given location $s \in \mathcal{S}$, we find its $k$ nearest landmarks and assign to it a landmark category $l \coloneqq \ell(s)$ as the most common landmark category among the $k$ landmarks.
Thus, the function
\[
    \ell(s): \mathcal S \rightarrow \mathscr{L}
\]
serves a labeling mechanism that maps each location within the city $\mathcal{S}$ to a corresponding landmark category in the set $\mathscr{L}$. 
Locations sharing the same landmark category (\textit{e.g.}, $l$) are grouped to form the functional zone $\mathcal{S}_l$, and we have $\mathcal{S} = \cup_{l \in \mathscr{L}} \mathcal{S}_l$. The left panel in Fig~\ref{fig:join-crime-landmark} visualizes the partition of urban functional zones in Valencia.
% (an overlay of a square grid covering the Valencia city area is created for the ease of visualization, with each grid cell assigned to a specific zone).

\subsection{Event mark definition}
\label{sec:event-mark-definition}

To accurately depict patterns of criminal activity across the city, it is crucial to consider contextual information about crime events, such as the type of crime and the environment setting in which it occurs. 
Currently, crimes are grouped by their crime types, for instance, \textit{Assault} (or \textit{Agresión}, in the original name). This categorization, however, may overlook important contextual differences. For instance, an assault near a restaurant and another near a bank are both categorized under \textit{Assault}, despite the distinct human activity patterns typical of dining and financial areas. By refining our crime categorization to account for these specific environment settings, we can enhance our understanding of crime dynamics.

We design a novel \textit{mark} associated with each event \citep{reinhart2018review} to categorize the crime events. 
The mark is designed to combine the event's crime category $c$ and the landmark category $\ell(s)$ of its location $s$, thus considering the urban functional zone that the event falls in. We denote the event mark as $c \times \ell(s)$.
For instance, as illustrated in the right panel of Fig~\ref{fig:join-crime-landmark}, the \textit{Assault} occurring in a restaurant zone on July 9th, 2016, is assigned the label $1 \times 6$ (representing \textit{Assault}$\times$ restaurant), while another \textit{Subtraction} on October 12th, 2018, in a financial zone receives the label $2\times 1$ (representing \textit{Subtraction}$\times$ financial).
The value space of the mark is a finite set $\mathscr{C} \times \mathscr{L}$ with a size of 21 (three crime categories and seven landmark categories).
We refer to the mark ``crime-landmark label'' of the event in the later discussion to reveal its practical meaning.
As we can see, this new mark derives a comprehensive categorization of crime events by including contexts of the observed event. Meanwhile, it allows for a detailed examination of crime patterns across different urban functional zones by analyzing incidents through the lens of their specific crime-landmark labels.

\subsection{Event dependence through multiple spaces}
\label{sec:multiple-network-space}

Crime events are ordered in time, and historical events will impact the probability, timing, or characteristics of future events \citep{mohler2011self, loeffler2018gun}. Such an impact is referred to as \textit{event dependence}.
To model the dependence among temporal events, we consider their relations over the geographic space with an underlying street network structure and the mark space (crime-landmark labels) characterized by an interaction network.

\begin{figure}[!t]
%\vspace{-.15in}
\centering
\begin{subfigure}[h]{.9\linewidth}
\includegraphics[width=\linewidth]{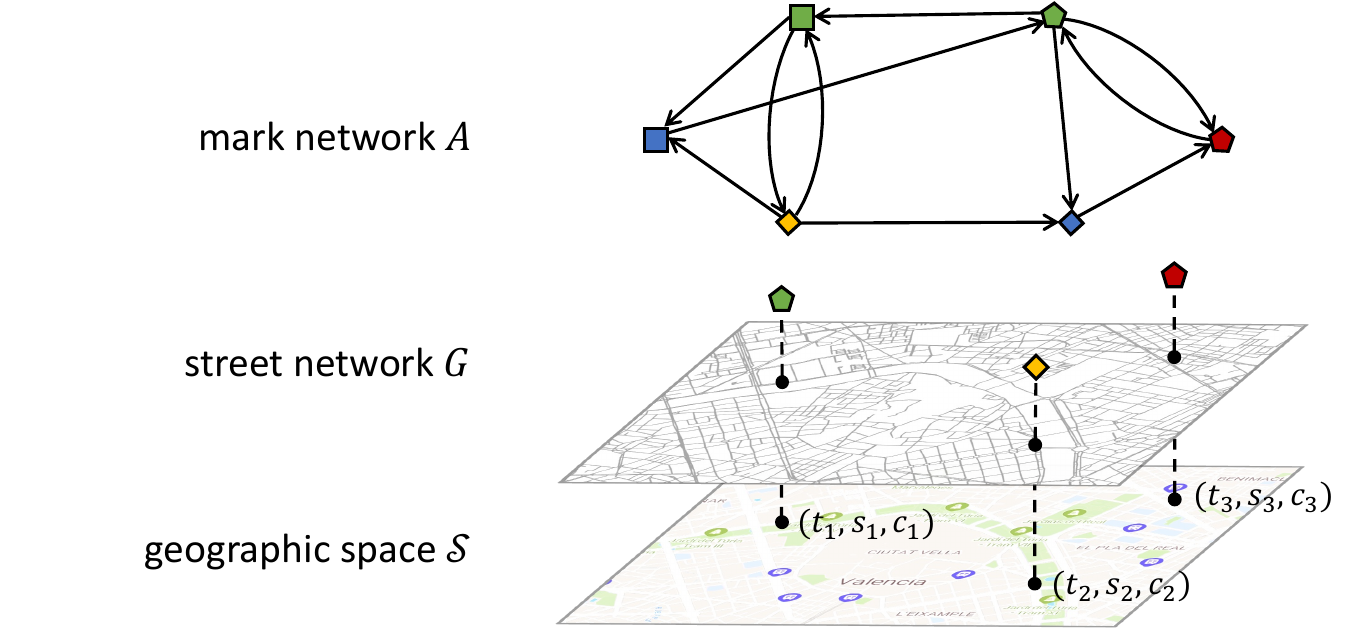}
\end{subfigure}
\caption{Multiple spaces for event dependence. An overlay of street network $G$ on top of the two-dimensional geographic space $\mathcal{S}$ is extracted using the real road information in Valencia for modeling the spatial connectivity of crime events. 
The crime events (black dots) can be mapped onto the corresponding edges of network $G$ according to their locations in the space $\mathcal{S}$.
Another network $A$ captures the event dependence over the mark space (marks represented by various shapes and colors), which is learned by the proposed model in Section~\ref{sec:point-process-modeling}. 
The multiple networks jointly depict the complex, multi-modal crime relation over the space of time, location, and event marks.
}
\label{fig:multiple-networks}
\vspace{-0.1in}
\end{figure}

To model the spatial relationship between crime events on the urban streets of Valencia, we overlay a street network structure $G$ on top of the continuous geographic space $\mathcal{S}$. This street network is constructed using the data from OpenStreetMap database \citep{OpenStreetMap}. The streets in Valencia are represented as linear segments linked at their endpoints. Note that the endpoints of these segments do not necessarily align with actual street intersections; they can be located in the middle of a street, such as on a curved street divided into multiple segments. These endpoints are treated as the nodes of the street network, while the street segments become the edges connecting these nodes. Each network edge is associated with an attribute, known as the \textit{edge weight}, indicating the length of the corresponding street segment measured in kilometers. 
The network is processed to be undirected to reflect the mobility patterns in street crimes in Valencia, where perpetrators commonly travel on foot or by bike in either direction along the streets \citep{bounceValencia}.
The street network consists of 8,043 nodes and 12,309 weighted, undirected edges, covering the entire city area of Valencia. 
It is worth noting that crime events are integrated into this network by being mapped to random locations on the network edges based on their geographic coordinates rather than being assigned to specific nodes.
Two layers at the bottom in Fig~\ref{fig:multiple-networks} illustrate such an overlay of the street network $G$ and the mapping of the crime events to the network edges.

Understanding the relation between event marks also provides valuable insights into characterizing the dependencies of crime events. 
By analyzing the sequence of observed marks, we can determine if certain events tend to be triggered by others in a specific pattern.
In this study, such dependencies are represented through a mark network, denoted as $A$. Each node of the mark network represents a distinct crime-landmark label (total of 21 nodes), and the events are assigned to the corresponding nodes based on their crime-landmark labels. 
The edges between these nodes indicate the potential relation between the crime-landmark labels they connect with. Such a relation can be directional, \textit{i.e.}, an observed crime with label $c\times \ell(s)$ may influence the occurrence of a future event with label $c'\times \ell(s')$ but not vice versa. Hence, the edges of the mark network are directional. 
Unlike the street network, which is derived directly from available geographic data, the mark network is established by learning a point process model detailed in the next section from the crime data. This model learns from the crime data to establish the directed and weighted edges of the mark network, indicating both the direction and strength of the dependencies between different event marks.
An example of the mark network is presented at the top of Fig~\ref{fig:multiple-networks}, highlighting directional relations among various event marks (crime-landmark labels).

\section{Point process modeling for event dependence}
\label{sec:point-process-modeling}

With the introduced event marks in Section~\ref{sec:data-processing}, we re-denote the processed data of $n$ observed crime events in \eqref{eq:raw-data} as
\[
    (t_1, s_1, c_1\times l_1), (t_2, s_2, c_2\times l_2), \dots, (t_n, s_n, c_n\times l_n),
\]
where $0 \leq t_1 < t_2 < \cdots < t_n \leq T, s_i \in \mathcal{S}, c_i \in \mathscr{C}$, and $l_i \coloneqq \ell(s_i) \in \mathscr{L}$. In the following, we present our point process modeling for understanding the multi-modal dependencies among the reported crime events over the street network.

% \zheng{Do we need to change the location space from $\mathcal{S}$ to $G$?}

\subsection{Spatio-temporal-network point processes}

% Crime events always exhibit the near-repeat phenomenon \citep{he2020discovering, townsley2003infectious}, as the occurrence of a crime will increase the likelihood of future crimes occurring in the nearby vicinity. This fact becomes the motivation for the adoption of self-exciting spatio-temporal point processes \citep{moller2003statistical, reinhart2018review} in this study, for their capability to model the contagious effect of crime events \citep{mohler2011self}.
Self-exciting spatio-temporal point processes \citep{moller2003statistical, reinhart2018review} are widely used in crime modeling to capture the contagious nature of crime events \citep{mohler2011self}.
Let $\mathcal{H}_t = \{(t_i, s_i, c_i\times l_i) \in \mathcal{H}_T|t_i < t\}$ denote the observed crime events happened before time $t$; we adopt a \textit{conditional intensity function} for each event category $c\times l$ to suggest the possibility of observing a new event with label $c\times l$ conditioning on the history. Specifically, the conditional intensity function at time $t$ and location $s$ is defined as
\begin{equation*}
    \lambda_{cl}\left(t, s \mid \mathcal{H}_t\right) =  \lim_{\Delta t \downarrow 0, \Delta s \downarrow 0} \frac{\mathbb{E}\left[\mathbb{N}_{cl}([t, t+\Delta t] \times B(s, \Delta s)) \mid \mathcal{H}_t\right]}{|B(s, \Delta s)| \Delta t}, \quad s \in \mathcal{S}_l,
\end{equation*}
where $B(s, \Delta s)$ is a ball centered at location $s$ with radius $\Delta s$. The $\mathbb{N}_{cl}$ is the counting measure for events with label $c\times l$, \textit{i.e.}, $\mathbb{N}_{cl}(A)$ is defined as the number of events with label $c\times l$ occurring within any subset $A \subseteq [0, T] \times \mathcal{S}$. This function essentially measures the rate at which events are expected to occur at a specific time and place based on historical data, with $\lambda_{cl}\left(t, s | \mathcal{H}_t\right) \geq 0$ for any arbitrary $c$, $l$, $t$ and $s$. To simplify the notation, we omit the $\times$ between $c$ and $l$ in the subscript.

Hawkes processes proposed in \citep{hawkes1971spectra} provide the self-exciting model formulation for capturing the triggering effects among events. It assumes that the occurrences of future events are positively influenced by the observed history, and the influence of past events is linearly additive. In this study, we model the conditional intensity function as follows:
\begin{equation}
    \lambda_{cl}(t, s \mid \mathcal{H}_t) = \mu_{cl} + \sum_{(t', s', c'\times l')\in \mathcal{H}_t}k(t', t, s', s, c'\times l', c\times l), \quad s \in \mathcal{S}_l.
    \label{eq:intensity-with-kernel}
\end{equation}
Here, $\mu_{cl}$ is a constant representing the base intensity of events with label $c \times l$. The $k$ function is the so-called influence kernel that captures the influence of a past incident $(t', s', c'\times l')$ on a current event $(t, s, c\times l)$.
This formulation allows for characterizing the influence of historical events on the likelihood of future events within the framework of the Hawkes process.
% Commonly the kernel function is assumed to be \textit{stationary}, that is, $k$ only depends on $t-t^\prime$ and $s-s^\prime$, which limits the model expressivity. In this work, we aim to capture complicated non-stationarity in spatio-temporal event dependencies by leveraging the strong approximation power of neural networks in kernel fitting. 

A separable form of the influence kernel has been commonly assumed in previous literature \citep{dong2023non, mohler2014marked, reinhart2018review, reinhart2018self, zhu2022spatiotemporal}.
% The modeling of the influence kernel over the multi-dimensional space can be complex and intractable. In our case, the $k$ has an input from a product space
The influence kernel $k$ can be expressed by the product of three individual kernel functions as
\[
    k(t', t, s', s, c'\times l', c\times l) = f(t', t)\cdot g(s', s) \cdot h(c'\times l', c\times l).
\]
The kernel functions $f, g, h$ characterize the event influence over the space of times, locations, and event marks, respectively.
We note that the separable form of the influence kernel enables a computationally efficient procedure for model fitting, given the large size of the data set. Meanwhile, the separable influence kernel can also provide us with interpretable results, as illustrated in Section~\ref{sec:results}.
In the following, we introduce the construction of these kernel functions in our context of modeling the street crime events within an urban environment. 

\paragraph*{Temporal kernel} We choose our temporal kernel $f$ to be an exponential function
\[
    f(t', t) = \beta e^{-\beta(t-t')}, \quad t>t'.
\]
Such a kernel function assumes the influence of a past event becomes significant in the near future and decays over time exponentially with a decaying rate $\beta > 0$, for subsequent incidents usually aggregate in time, occurring sooner after previous crimes.

\begin{wrapfigure}{r}{.45\linewidth}
\vspace{-.2in}
\centering
\begin{subfigure}[h]{.85\linewidth}
\includegraphics[width=\linewidth]{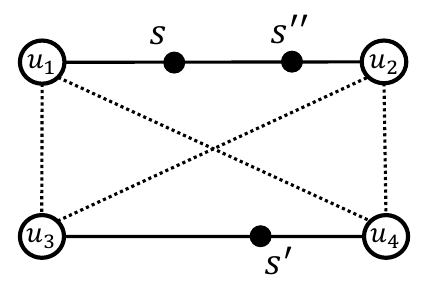}
\end{subfigure}
\vspace{-0.1in}
\caption{Two scenarios for calculating the street network distance: (i) When two locations (\textit{e.g.}, $s$ and $s'$) are on different edges (the solid lines), their network distance depends on the lengths of four shortest paths between their adjacent nodes (the dashed lines). (ii) When two locations are on the same edge (\textit{e.g.}, $s$ and $s''$), their network distance is the straight-line (Euclidean) distance between them.
% The $d_{\text{net}}(s, s')$ equals to the length of the path $s\to u_2 \to u_3 \to s'$, and the $d_{\text{net}}(s, s'')$ equals to the length of the path $s \to s''$.
}
\label{fig:network-distance}
\vspace{-0.15in}
\end{wrapfigure}

\paragraph*{Street-network-based spatial kernel} In our case, criminal activities appear on the city street network, and criminals typically use roads to flee crime scenes rather than traveling in straight lines, which is impractical due to urban structures, such as buildings. Therefore, the Euclidean distance between event locations becomes unsuitable for assessing the spatial connectivity between crime events.
% Thus, adopting a network-based distance measurement is crucial for accurately modeling criminals' movement patterns and analyzing the dispersion of crime events across urban areas.
% an Euclidean-distance-based spatial kernel may fail to recover the underlying patterns of the contagious effect of historical crimes. 
% Figure~\ref{fig:empirical-kernel} visualizes the empirical triggering effect of crime incidents with their distance measured over the Euclidean space and the street network. The incident effect measured based on the network distance shows a distinguishable decaying pattern with the increase of time lag or distance, while the effect measured using the Euclidean distance is noisy without exhibiting interpretable patterns.
Favored by the overlay of the street network, we adopt a street network distance \citep{wei2020distance}, denoted as $d_{\rm net}(s, s')$, for calculating the travel distance between any two locations $s$ and $s'$ on the network edges.
% based on the shortest-path distance among network nodes. 
The calculation of $d_{\rm net}$ involves two scenarios, as illustrated in Fig~\ref{fig:network-distance}: (i) The movement from $s$ to $s'$ on different edges involves moving from $s$ to an adjacent node ($u_1$ or $u_2$), traversing the shortest path (indicated by dashed lines in Fig~\ref{fig:network-distance}) to a node ($u_3$ or $u_4$) on the edge that $s'$ falls on, and finally proceeding to $s'$. There are four possible paths between $s$ and $s'$: $s \to u_1 \mathrel{\leadsto} u_3 \to s'$, $s \to u_1 \mathrel{\leadsto} u_4 \to s'$, $s \to u_2 \mathrel{\leadsto} u_3 \to s'$, and $s \to u_2 \mathrel{\leadsto} u_4 \to s'$, where the $\mathrel{\leadsto}$ represents the shortest path over the network between two nodes. Then, $d_{\rm net}(s, s')$ equals the shortest length of these four paths; (ii) For $s$ and $s''$ on the same edge, $d_{\rm net}(s, s'')$ is simply the straight-line (Euclidean) distance between them.
Based on the street network distance, we propose a Gaussian spatial kernel, defined as
\[
    g(s', s) = \frac{1}{2\pi \sigma^2}e^{-\frac{d_{\rm net}^2(s, s^\prime)}{2\sigma^2}}.
\]
This kernel function indicates that the influence of an event decays as the distance increases. Parameter $\sigma>0$ determines the scale of influence across the street network, illustrating how spatial interactions diminish over distance.
% aligning with those displayed by the empirical kernel.

% \begin{figure}[!tb]
% %\vspace{-.15in}
% \centering
% \begin{subfigure}[h]{.23\linewidth}
% \includegraphics[width=\linewidth]{empirical_kernel_ED_1.pdf}
% % \caption{Vanilla Hawkes}
% \end{subfigure}
% \begin{subfigure}[h]{.23\linewidth}
% \includegraphics[width=\linewidth]{empirical_kernel_ED_2.pdf}
% % \caption{Vanilla Hawkes}
% \end{subfigure}
% \hspace{0.1in}
% \begin{subfigure}[h]{.23\linewidth}
% \includegraphics[width=\linewidth]{empirical_kernel_NWD_1.pdf}
% % \caption{Vanilla Hawkes}
% \end{subfigure}
% \begin{subfigure}[h]{.23\linewidth}
% \includegraphics[width=\linewidth]{empirical_kernel_NWD_2.pdf}
% % \caption{Vanilla Hawkes}
% \end{subfigure}
% \caption{Empirical kernels based on Euclidean distance and street-network-based distance.
% }
% \label{fig:empirical-kernel}
% % \vspace{-0.1in}
% \end{figure}

% $t \in [0, T]$ represents the time of the current event, $s \in \mathcal{S}$ represents its location, and $cl \in [D]$ represents the index of its crime-landmark type. Similarly, $t' \in [0, T]$, $s' \in \mathcal{S}$, and $cl' \in [D]$ represent the time, location, and index of the crime-landmark type, respectively, for the historic event. We use a spatio-temporal kernel $g$ to capture the impact of the historic event on the current event over time and space, and a coefficient $\alpha_{cl, cl^\prime}$ to measure the magnitude of the triggering effect from an past event with label $cl^\prime$ to the current event with label $cl$. Lastly, the $\mu_{cl}$ represents the background intensity of events with label $cl$. 

\paragraph*{Interactions between event marks} 
To model the interactions between event marks that are categorical, we represent the kernel function $h$ using a set of coefficients $\{\alpha_{cl, c'l'}\}_{c, c' \in \mathscr{C}, l, l' \in \mathscr{L}}$, where 
\[
    h(c'\times l', c\times l) = \alpha_{cl, c'l'}
\]
captures the influence of a historical event with mark $c'\times l'$ on a future event with mark $c\times l$.
A larger value of $\alpha_{cl, c'l'}$ contributes more to the conditional intensity function, suggesting a higher possibility of observing a future event marked by $c\times l$ given an observed event mark $c'\times l'$.
% representing a stronger dependence between these two event marks.
Note that such an interaction can be directional, that is, $\alpha_{cl, c'l'} \neq \alpha_{c'l', cl}$. 
All the coefficients are set to be non-negative, and a zero-value $\alpha_{cl, c'l'}$ means no influence from events with mark $c'\times l'$ to events with mark $c\times l$.
The mark network $A$ is established accordingly from the coefficients. When $\alpha_{cl, c'l'} > 0$, a directed edge from the node representing crime-landmark label $c'\times l'$ to the node representing label $c\times l$ is created, with the edge weight assigned as $\alpha_{cl, c'l'}$.
% \zheng{write something about the connection between coefficients and the mark network}

\vspace{0.2in}
Following the chosen kernel functions, the conditional intensity function for a crime event with mark $c\times l$ at time $t$ and location $s$ is modeled as follows

\begin{equation}
    \begin{aligned}
        \lambda_{cl}(t, s) 
        &= \mu_{cl} + \sum_{(t', s', c'\times l') \in \mathcal{H}_t}\alpha_{cl, c'l'}\beta e^{-\beta(t-t^\prime)}\frac{e^{-\frac{d_{\rm net}^2(s, s^\prime)}{2\sigma^2}}}{2\pi \sigma^2}, \quad s \in \mathcal{S}_l.
        % e^{-\frac{d^2(s^\prime, s_l)}{2\sigma_l^2}} 
    \end{aligned}
    \label{eq:proposed-intensity}
\end{equation}
% The function $\mu_c(\cdot)$ with regard to different landmark labels (determined by the incident location $s$) models the background intensity of type-$c$ crime in different urban functional regions. 
The base intensity $\mu_{cl}$ is estimated from the data.
The influence kernel is chosen to integrate to $\alpha_{cl, c'l'}$, providing a natural interpretation of the coefficient:  $\alpha_{cl, c'l'}$ is the expected number of crime events with mark $c\times l$ triggered by an observed event with mark $c'\times l'$.
Here, for notation simplicity, we omit the dependence on history $\mathcal{H}_t$ in the intensity function and use common shorthand $\lambda_{cl}(t, s)$ to denote $\lambda_{cl}(t, s\mid \mathcal{H}_t)$.
Note that it is possible to allow different spatial and temporal decays for events with different crime landmark labels. Yet, this approach would significantly increase the number of model parameters.

\subsection{Influence kernel learning with graph neural networks}
\label{sec:gnn-learning-kernel}

The learning of the coefficients $\{\alpha_{cl, c'l'}\}_{c, c' \in \mathscr{C}, l, l' \in \mathscr{L}}$ plays an essential role in understanding the mark interactions and the characterization of the event dynamics.
% the event dependence over the mark space is captured by the coefficients $\{\alpha_{cl, c'l'}\}_{c, c' \in \mathscr{C}, l, l' \in \mathscr{L}}$, which indicate the mark network structure 
% Each coefficient indicates the contribution of a certain mark to the conditional intensity of events with another mark. 
Traditionally, these coefficients have been directly estimated from data, as outlined in various studies \citep{mohler2014marked, reinhart2018self, zhu2022spatiotemporal}. However, recent advancements in point process models have showcased the value of incorporating prior knowledge of event marks, known as \textit{features}, into the modeling of these coefficients and the mark interactions.
For example, when modeling the interactions between different social media users (marks), the work of Group Network Hawkes Process \citep{fang2023group} treats these users as network nodes and leverages their characteristics (the features of the marks) to effectively identify the group interactions and influential users in social networks. 
In our case, the event marks are defined by the combinations of multiple crime and landmark categories. It is reasonable to believe that marks sharing the same crime or landmark category tend to exhibit stronger interactions than those with differing categories.
Therefore, these crime or landmark categories that compose the mark can be regarded as the mark features that we can leverage in the coefficient modeling.
% \zheng{maybe some data analysis}
% Leveraging prior knowledge about the marks allows for a more comprehensive and effective understanding of event dynamics.
% can offer a comprehensive insight into the complex event relations, allowing for understanding the dynamics of event occurrences more effectively.

We introduce a novel approach via GNNs to model the coefficients, leveraging their ability to integrate nodal features in learning node similarity.
We first decompose the coefficients into two components as follows:
\[
    \alpha_{cl, c'l'} = a_{cl, c'l'} \cdot p_{cl, c'l'},
\]
where $a_{cl, c'l'} > 0$ and $0 \leq p_{cl, c'l'} \leq 1$ are both scalars.
Together, these two components can be viewed as the \textit{strength} and the \textit{chance} of the interaction between two marks. The term $p_{cl, c'l'}$, modeled by a GNN, will incorporate the mark features and capture the graph topology by providing the likelihood for any $c \times l$ to have a connection to the rest of $c' \times l'$; meanwhile, the $a_{cl, c'l'}$ captures the weights on the edges in the mark network, indicating the strength of the connection. 

We model these two components separately. For the \textit{strength} $a_{cl, c'l'}$, we treat it as a trainable scalar that is learned from data. For the modeling of the \textit{chance} $p_{cl, c'l'}$, we use Graph Attention Networks (GAT) \citep{velickovic2018graph} to take the mark features into account.
The feature of mark $c\times l$ can be denoted by a column vector $X_{cl} \in \mathbb{R}^{D}$, which is the concatenation of the one-hot vectors of the crime category $c$ and the landmark category $l$.
These feature vectors are passed through the GAT to compute attention scores between pairs of marks. Each score quantifies the likelihood that mark $c'\times l'$ influences mark $c\times l$, based on their feature vectors. The scores are obtained using a multi-head self-attention mechanism over graphs with $R$ attention heads. In the $r$-th attention head, the score is
\[
    e^r_{cl,c'l'} = \mathrm{LeakyReLU}\!\left( {\bm{b}^r}^\top \big[\, \bm{W}^r X_{cl} \,\|\, \bm{W}^r X_{c'l'} \,\big] \right),
\]
where $\bm{W}^r \in \mathbb{R}^{D'\times D}$ is the shared linear transformation for each mark feature, $\bm{b}^r \in \mathbb{R}^{2D'}$ is a learnable vector, and $\|$ denotes concatenation. The Leaky ReLU nonlinearity \citep{maas2013rectifier} is defined as
\[
    \mathrm{LeakyReLU}(x) = \max(0, x) + b \min(0, x), \quad b=0.2,
\]
consistent with the original GAT implementation \citep{velickovic2018graph}. For a fixed target mark $c\times l$, the attention scores $\{e^r_{cl,c'l'}\}$ are normalized across all possible $c'\times l'$ using the softmax function to yield the attention-based interaction probability from the $r$-th head:
\begin{equation}
    p^r_{cl,c'l'} = 
    \frac{\exp\!\left( e^r_{cl,c'l'} \right)}
    {\sum\limits_{c'' \in \mathscr{C},\, l'' \in \mathscr{L}} \exp\!\left( e^r_{cl,c''l''} \right)}.
    \label{eq:normalized-attention-weights}
\end{equation}
Finally, the interaction probability $p_{cl,c'l'}$ is obtained by averaging over all $R$ heads:
\[
    p_{cl,c'l'} = \frac{1}{R}\sum_{r=1}^R p^r_{cl,c'l'}.
\]

Note that GAT ensures $\sum_{c' \in \mathscr{C}, l' \in \mathscr{L}}p_{cl, c'l'} = 1$, that is, the $p_{cl, c'l'}$ collectively form a probability distribution over possible source marks for each target mark $c\times l$.
The hyper-parameter to be determined in advance is the number of attention heads $R$ to achieve the balance between model flexibility and generability. The learnable parameters are $\{\bm{b}^r, \bm{W}^r\}_{r=1}^{R}$ in GAT and the interaction strength $\{a_{cl, c'l'}\}_{c, c' \in \mathscr{C}, l, l' \in \mathscr{L}}$.

\section{Model estimation}
\label{sec:model-estimation}

We now discuss the estimation of model parameters based on the Maximum Likelihood Estimation (MLE) approach \citep{reinhart2018review}. The units for measuring the event time and distance are days and kilometers, respectively, throughout the model estimation and empirical experiments.

We first estimate the base intensity $\{\mu_{cl}\}_{c\in \mathscr{C}, l\in \mathscr{L}}$ as the average number of observed events with mark $c \times l$ per space-time unit (\textit{i.e.}, per kilometer per day) divided by a constant, which serves as a hyperparameter to adjust the baseline intensity and can be selected via cross-validation. In our experiments, perform 4-fold cross-validation on the training set and select 50 from the candidate set $\{1, 2, 5, 10, 20, 50, 100\}$. We observe that an 
overestimation of the base intensity (e.g., using a dividing constant of 1) will suppress the learned triggering effects, causing the model to underestimate event dependencies and degrade in predictive performance. In practice, this constant value can be chosen by cross-validation or informed by domain knowledge.
Other non-parametric procedures for estimating the base intensity using stochastic declustering \citep{mohler2011self, zhuang2019semiparametric} or kernel density estimation \citep{mohler2014marked, reinhart2018self, yuan2019multivariate} have been adopted in previous literature on modeling self-exciting crime events. Compared with these methods, our approach provides a more computationally efficient procedure, particularly for large-scale crime data sets (\textit{e.g.}, more than 10,000 crimes) \citep{reinhart2018review}, and avoids the model identification issue when Gaussian kernels are used in both base intensity and influence kernel \citep{reinhart2018self}. 
Additional results are provided in Appendix~\ref{app:non-para-base-estimation}, demonstrating the estimation accuracy and computational benefits of our method compared with traditional stochastic declustering.
By estimating base intensities for various crime types and urban functional regions, our approach also captures the heterogeneity in event occurrence across both geographic space and mark space.

The influence kernel is estimated by maximizing the log-likelihood function of the point process model \citep{daley2003introduction}. We denote the parameters in the influence kernel as $\theta \coloneqq \{\{\bm{b}^r, \bm{W}^r\}_{r=1}^{R}, \{a_{cl, c'l'}\}_{c, c' \in \mathscr{C}, l, l' \in \mathscr{L}}, \beta, \sigma\}$. The log-likelihood function of observing $\mathcal{H}_T = \{(t_i, s_i, c_i\times l_i)\}_{i=1}^{n}$ on $[0, T] \times \mathcal{S}$ is given by
\begin{equation}
    \begin{aligned}
        L(\theta)&=\sum_{i=1}^n \log \lambda_{c_il_i}\left(t_i, s_i\right)-\sum_{c\in\mathscr{C}, l\in \mathscr{L}}\int_0^T \int_{\mathcal{S}} \lambda_{cl}(t, s) ds dt
        % \\&=\sum_{i=1}^{n}\log \left(\mu_{c_il_i} + \sum_{j=1}^{i-1}\alpha_{c_il_i, c_jl_j}\beta e^{-\beta(t_i-t_j)}\frac{e^{-\frac{d^2_{\rm net}(s_i, s_j)}{2\sigma^2}}}{2\pi \sigma^2} \right)
        ,
    \end{aligned}
    \label{eq:pp-log-likelihood}
\end{equation}
where $\theta$ is incorporated into the conditional intensity function (see Appendix~\ref{app:derivation-calculation-llk} for log-likelihood derivation).
Due to the existence of graph neural networks in our model and the large data size, solving the M-step in the classic expectation-maximization (EM) algorithm for point processes \citep{liu2021point, veen2008estimation, zhu2022spatiotemporal} becomes intractable and overwhelming. Therefore, we adopt the commonly-used optimization strategy of stochastic gradient descent \citep{robbins1951stochastic} to estimate the model parameters $\theta$.
The crime data set used for model training is separated into multiple event sequences by consecutive fixed-length time windows. The obtained event sequences will be retrieved in random order with a fixed batch size. Each retrieved batch of the event sequences is used to compute the gradient of the loss function with regard to the model parameters using backpropagation \citep{rumelhart1986learning}. The model parameters are then updated along the computed gradient with a chosen learning rate $\eta$.
In our case, the loss function for each batch is the summation of the negative log-likelihoods $-L(\theta)$ of all the sequences in that batch.  Algorithm~\ref{alg:learning-model-parameters} summarizes the learning procedure for the parameters $\theta$, where we set the batch size $M = 3$, learning rate $\eta = 1.0$, and epoch number $E=1,500$ in our experiments.
The validity of using multiple subsequences for learning the parameters can be guaranteed by setting the length of the time window used for splitting the entire sequence (for example, 120 days) much larger than the scale of the decaying temporal effect of historical events (around 30.77 days by the final learned model).

\begin{algorithm}[!t]
\begin{algorithmic}
    \STATE {\bfseries Input}: Training set $\{\mathcal{H}^j_T\}_{j=1}^{J}$ with $J$ non-overlapping subsequences, where $\mathcal{H}_T^j = \{(t_i, s_i, c_i\times l_i)\}_{i=1}^{n_j}$ and $\cup_{j=1}^{J}\mathcal{H}^j_T=\mathcal{H}_T$; batch size $M$; epoch number $E$; learning rate $\eta$. \;
    \STATE {\bfseries Initialization:} model parameters $\theta^{(0)}$, first epoch $e=0$.\; 
    \WHILE{$e < E$}
        \FOR{each batch $\{\mathcal{H}_T^{j_1}, \dots, \mathcal{H}_T^{j_M}\}$ with size $M$}
            \STATE 1. Compute the negative log-likelihood $-L^{j}(\theta^{(e)})$ using $\mathcal{H}_T^{j}$ for $j \in \{j_1, \dots, j_M\}$, according to \eqref{eq:pp-log-likelihood}. \;
            % \vspace{.05in}
            \STATE 2. Compute the gradient $\left. \bm{g}^{(e)} = \nabla_{\theta}\left(\sum_{j_1, \dots, j_M}-L^j(\theta)\right)\right \rvert_{\theta = \theta^{(e)}}$ using backpropagation. \;
            % \vspace{.05in}
            \STATE 3. Update the model parameters: $\theta^{(e+1)} \leftarrow \theta^{(e)} - \eta \bm{g}^{(e)}$. \;
            % \vspace{.05in}
        \ENDFOR
        
        $e \leftarrow e + 1$
    \ENDWHILE
    \RETURN Learned model parameter $\theta^{(E)}$.
\end{algorithmic}
\caption{Model parameter estimation using stochastic gradient descent}
\label{alg:learning-model-parameters}
\end{algorithm}

% \begin{remark}
% Proposition~\ref{prop:approximation-of-integral} leads to a computationally efficient calculation of the integral with complexity $\mathcal{O}(N)$. We denote the upper bound of the relative error as $\overline{\epsilon_2}$ and its dependence on hyper-parameters $A$ and $c$ is illustrated in Fig.~\ref{fig:upper-bound-surface}.
% In general, a larger $c$ results in a more expressive spatial kernel but requires a larger $A$ to control the approximation error. In practice, we select $c = 0.1$ and $A = 0.35$ to limit the relative error $\epsilon_2$ under 0.05 and ensure a certain level of expressiveness for the spatial kernel.
% \end{remark}
\vspace{0.1in}
\noindent \textit{Remark:} The computational cost of the loss function mainly lies in the evaluation of the first term in \eqref{eq:pp-log-likelihood}, which involves evaluations of the influence kernel between each pair of events in the event sequence. By dividing the entire training sequence with $n$ events into $J$ subsequences with each of $n_j$ events, the complexity of computing \eqref{eq:pp-log-likelihood} over the entire data set can be reduced from $\mathcal{O}(n^2)$ down to $\sum_{j=1}^{J}\mathcal{O}(n_j^2) \approx \mathcal{O}(n^2/J^2)$. In fact, we are eliminating the overwhelming and unnecessary evaluations of the influence kernel between event pairs that are far away enough over time so that the earlier event has little or no influence on the latter one. 
Fig~\ref{fig:tradeoff-num-subseq} in Appendix~\ref{app:additional-experiments} shows the model training time and the model's goodness-of-fit on the training data set with different $J$s. With a proper $J$, enhanced model computational efficiency can be attained without degrading the model performance. In our experiments, we choose $J=12$ to achieve a balance between model performance and computational efficiency (\textit{i.e.}, the length of the time window for each subsequence is 120 days).
% \zheng{add a set of results about: 1) model performance 2)computational cost with different lengths of subsequences.}

% \zheng{computational efficiency by cutting the entire sequence into pieces?}

% \subsection{Low-rank and sparsity regularization}

% % This part is important (so many parameters in practice and we need the regularization).

% The optimization problem with sparsity regularization can be expressed as following:

% \begin{equation}
%     \min_{\bm{A}, \mu, \beta, \sigma} \mathcal{L}(\bm{\theta}) = \ell(\bm{\theta}) + \lambda_2\|\bm{A}\|_{1},
%     \label{eq:loss-function}
% \end{equation}

% where $\bm{A}$ is the coefficient matrix.

% \subsection{Uncertainty quantification}

\section{Results}
\label{sec:results}

We now present the results by analyzing the crime data set in Valencia (Spain), and further demonstrate the competitive performance of our proposed model (referred to as \texttt{STNPP}) in predicting future crime rates and understanding the dynamics of crime events \footnote{Code available at \href{https://github.com/McDaniel7/Spatio-Temporal-Network-Point-Process}{https://github.com/McDaniel7/Spatio-Temporal-Network-Point-Process}}. The entire data set is partitioned into two parts. The first part includes data from 2015 through 2018, which is used to estimate the model parameters and evaluate the goodness-of-fit of the model. The second part contains data from 2019 and is used for assessing the model's predictive performance.

% \zheng{need a ablation study to show (i) the hyper-parameter selection; (ii) the impact of thresholding the attention weights in GAT when learning the graph topology.}

\subsection{Model validation}
\label{sec:model-validation}

We first validate our model from two aspects: the determination of the hyper-parameter $R$ and the goodness-of-fit of the chosen model on the crime data.

% \begin{table}[!t]
%   \caption{Model performance with different number $R$ of attention heads in 5-fold cross-validation.}
%   % \vspace{-0.1in}
%   \centering
%   \resizebox{.98\linewidth}{!}{
%   \begin{threeparttable}
%   \begin{tabular}{cccccc}
%     \toprule
%     \toprule
%     % & \multicolumn{3}{c}{Traffic congestion (5 nodes)}\\
%     % \cmidrule(lr){2-4} \cmidrule(lr){5-7}  \cmidrule(lr){8-10} \cmidrule(lr) {11-13}
%     Model & MAE (scarce) ($\downarrow$) & MAE (frequent) ($\downarrow$) & MAE (total) ($\downarrow$) & Training $\ell$ ($\uparrow$) & Testing $\ell$ ($\uparrow$) \\
%     \midrule
%     $R=2$ & 0.886 & 7.257 & 31.570 & $-2.652_{(0.007)}$ & -2.422 \\
%     $R=4$ & 0.884 & 6.395 & 30.279 & $-2.600_{(0.004)}$ & -2.387 \\
%     $R=6$ & 0.871 & 6.053 & \textbf{29.553} & $-2.556_{(0.005)}$ & -2.323 \\
%     $R=8$ & \textbf{0.830} & \textbf{5.758} & 29.554 & $-2.530_{(0.002)}$ & -2.288 \\
%     $R=12$ & 0.843 & 6.043 & 30.303 & $-2.532_{(0.004)}$ & \textbf{-2.287} \\
%     $R=16$ & 0.852 & 6.432 & 32.183 & $-2.518_{(0.005)}$ & -2.292 \\
%     \bottomrule
%     \bottomrule
%   \end{tabular}
%   % \begin{tablenotes}
%   % \item *Numbers in parentheses are standard errors for three independent runs.
%   % \end{tablenotes}
%   \end{threeparttable}
%   }
%   % \vspace{-0.2in}
%   \label{tab:number-of-attention-head}
% \end{table}

\begin{wrapfigure}{r}{.5\linewidth}
\vspace{-.25in}
\centering
\begin{subfigure}[h]{.98\linewidth}
\includegraphics[width=\linewidth]{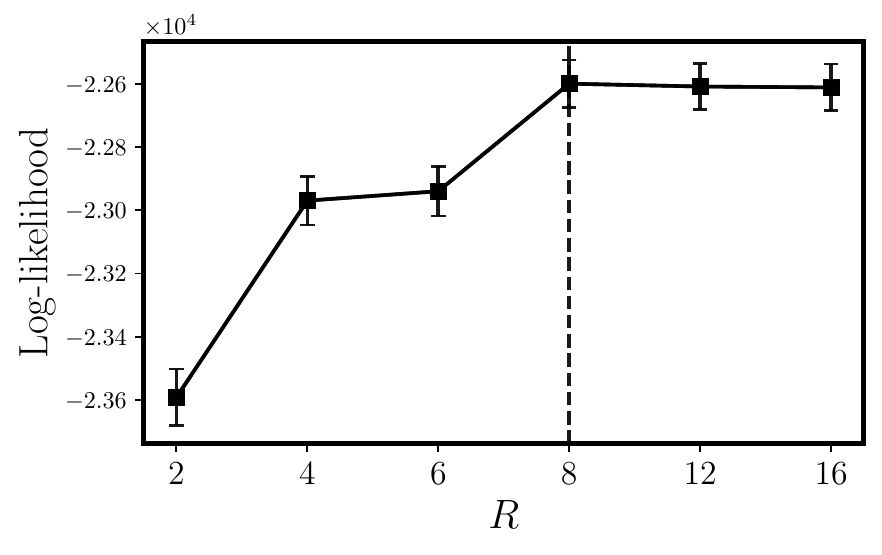}
\end{subfigure}
\vspace{-0.1in}
\caption{Cross-validation for selecting attention head number $R$. Results are averaged over four folds with standard deviations reported. The dashed line marks the optimal $R$.
}
\label{fig:cv}
\vspace{-0.15in}
\end{wrapfigure}

An appropriate choice of the number of attention heads $R$ in GAT needs to be determined in advance, which can be achieved using cross-validation. We first divide the training data from 2015 to 2018 into 12 subsequences with the same time window length of 120 days. Then, we adopt 4-fold cross-validation on the training data to determine the value of $R$. Given a choice of $R$, all the 12 subsequences are shuffled randomly and split into four groups. One round of cross-validation involves taking one group as the hold-out data, training the model with the remaining groups, and evaluating the trained model on the hold-out data. The final model performance is obtained by averaging the metrics over four independent rounds.
We compare the performance of the model with $R=2, 4, 6, 8, 12$, and $16$ attention heads in terms of the model log-likelihood on the hold-out data, which evaluates the model generalization ability to the unseen data. Better model performance is indicated by a higher hold-out log-likelihood.
Fig~\ref{fig:cv} reports the averaged hold-out log-likelihood with different attention heads. According to the results, we choose $R=8$ as an optimal choice in the remaining experiments in this study.

\begin{figure}[!t]
%\vspace{-.15in}
\centering
\begin{subfigure}[h]{.98\linewidth}
\includegraphics[width=\linewidth]{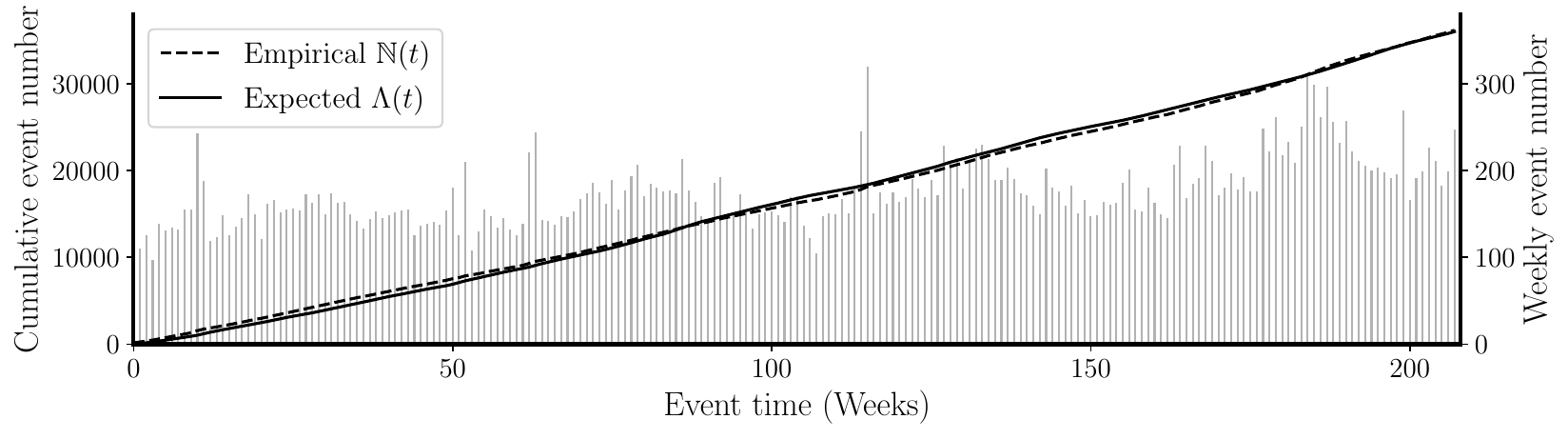}
\end{subfigure}
\caption{Empirical and expected cumulative number of events against the event times, represented by the black and red lines, respectively. The data period is from 2015 to 2018 (a total of 208 weeks). The grey vertical lines indicate the weekly number of crime events.
}
\label{fig:cumulative-event-num}
\vspace{-0.1in}
\end{figure}

Another model assumption -- the stationarity of the influence kernel needs to be validated by investigating the model's fit with the training data. Stationarity means that the model parameters do not vary over time, indicating that the pattern of event influence remains consistent. Previous research on crime modeling with point processes has frequently made this assumption, but often without adequately verifying its validity.
The work of the non-stationary ETAS model \citep{kumazawa2014nonstationary} presents a method to test the goodness-of-fit of a stationary point process model to the data by comparing the expected cumulative number of events computed from the learned model and empirical cumulative number of events. 
In our context, given the learned model $\hat{\lambda}_{cl}$, the expected cumulative number of events in the time interval $[0, t]$ is computed as $\Lambda(t) = \int^{t}_{0}\int_{\mathcal{S}}\sum_{c\in \mathscr{C},l\in \mathscr{L}}\hat{\lambda}_{cl}(t, s)dsdt$. If the model represents a good approximation of the real data, we expect that $\Lambda(t)$ and the empirical cumulative event counts $\mathbb{N}(t) = \sum_{c\in\mathscr{C}, l\in\mathscr{L}}\mathbb{N}_{cl}(t)$ are close. We fit the model using the entire training set, and plot the $\mathbb{N}(t)$ and $\Lambda(t)$ from 2015 to 2018 in Fig~\ref{fig:cumulative-event-num}. The consistent match between the empirical and expected cumulative event numbers suggests that the underlying data dynamics are stationary, and the assumption of kernel stationarity is reliable when fitting the data.

\subsection{Data fitting and in-sample estimations}

\begin{figure}[!t]
%\vspace{-.15in}
\centering
\begin{subfigure}[h]{.98\linewidth}
\includegraphics[width=\linewidth]{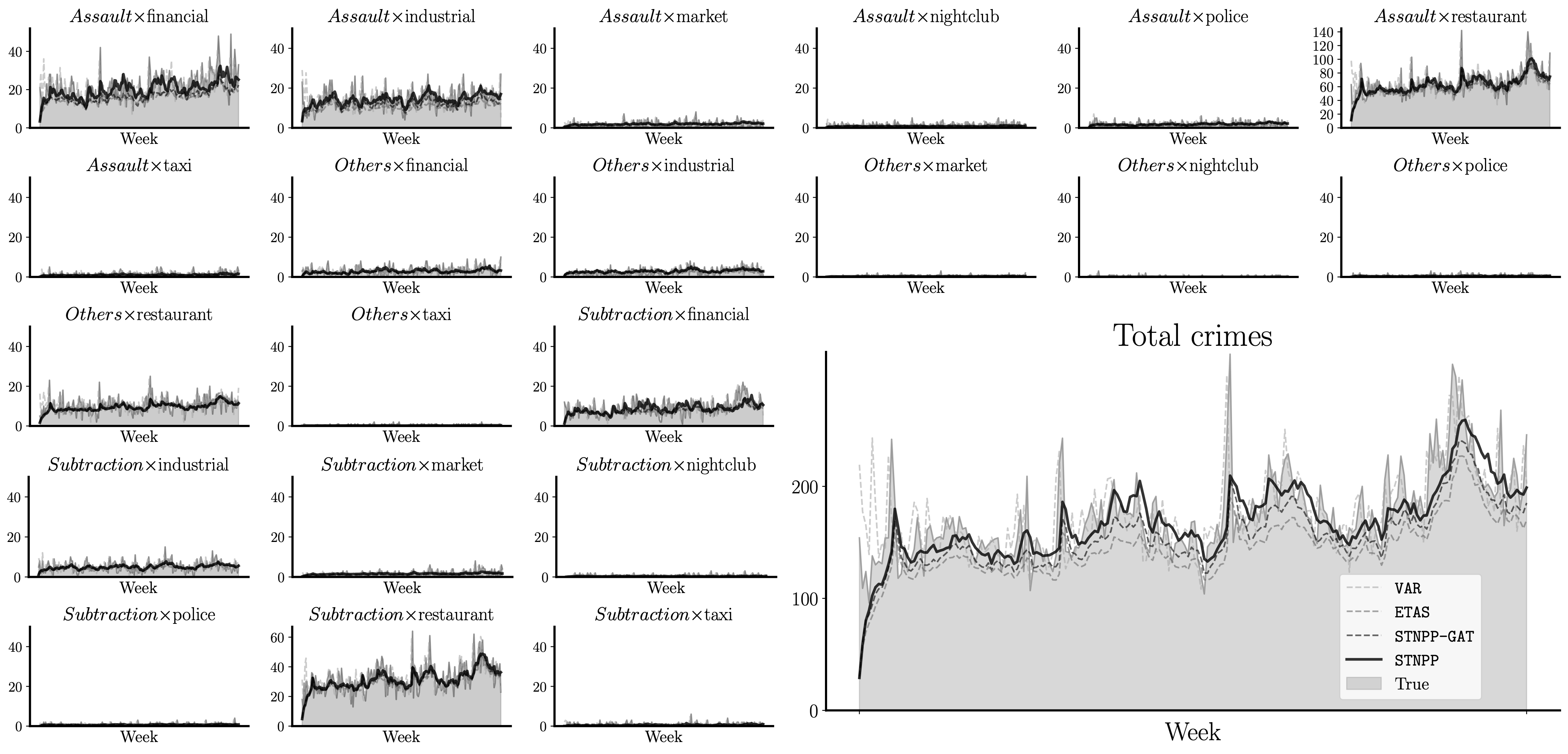}
\end{subfigure}
\caption{In-sample estimation of the number of crime events from 2015 to 2018 by different models. The red lines represent the in-sample estimations by our model \texttt{STNPP}. The dashed blue, yellow, and green lines represent the in-sample estimations by three baselines. The grey areas indicate the number of true observations.
}
\label{fig:in-sample-estimation}
\vspace{-0.1in}
\end{figure}

We then fit the model on the entire training data from 2015 to 2018 and analyze the results. To quantitatively demonstrate the effectiveness of our model, we compare our model with different baselines in terms of the fitted log-likelihood on the training data, the Akaike Information Criterion (AIC) \citep{akaike1974new, akaike1998information} of the model, and the mean absolute error (MAE) of the in-sample estimation of the number of the crime events. The log-likelihood, computed by \eqref{eq:pp-log-likelihood} using training data, measures the model goodness-of-fit to the training data. The AIC considers both the model fit to the data and the model complexity. It is described as $AIC = -2 \max_{\theta}\log L(\theta) + 2k$, where $\log L(\theta)$ is the model log-likelihood and $k$ is the number of model parameters to be estimated. The in-sample estimation of the event number over a given time interval can be performed as follows: we fit the model using the entire training data, feed the same data into the fitted model, and calculate the integral of the conditional intensity function over the time interval as the estimated number of events.
In practice, the in-sample estimation of number of events with mark $c\times l$ over $[t_1, t_2]$ can be calculated by $\int_{t_1}^{t_2}\int_{\mathcal{S}}\hat{\lambda}_{cl}(t, s)dsdt$.
We evaluate the in-sample estimation of event numbers during each week using our model \texttt{STNPP}, and compare its performance with four baselines, including two predictive time series models, one point process model, and an ablated variant of our model: 
(1) The persistence forecast (\texttt{Persistent}) that uses the event number in the previous week as the estimation; 
(2) Vector autoregression (\texttt{VAR}), which is a statistical model for analyzing and predicting multivariate time series data; 
(3) Epidemic-type aftershock sequence (\texttt{ETAS}) model \citep{ogata1998space} with a diffusion-type kernel using Euclidean distance; 
% (3) The \texttt{STNPP} but using Euclidean distance and without GAT (\texttt{STNPP-ND-GAT});
(4) The \texttt{STNPP} without GAT (\texttt{STNPP-GAT}).
We slightly modify the ETAS model by incorporating a set of coefficients to account for the interactions between different event marks, since the original ETAS model cannot deal with multiple event types (see Appendix~\ref{app:additional-experiments} for details). 
Fig~\ref{fig:in-sample-estimation} visualizes the in-sample estimations by different models on the number of each event type and the total events from 2015 to 2018, alongside the actual observed values.
Our model effectively recovers both the overall temporal trend in total event numbers and the specific temporal patterns for event types that occurred either frequently or infrequently during the training period. Note that such heterogeneity in the event dynamics is simultaneously captured by a holistic model instead of fitting independent models for each type of event.
% Our model can not only recover the temporal trend of the total event numbers but also simultaneously capture the temporal dynamics of the number of events with those marks that have either frequent or rare occurrences during the training period.

\begin{table}[!t]
  \caption{Quantitative results of data fitting and in-sample estimation. Bold indicates the best performance.}
  % \vspace{-0.1in}
  \centering
  \resizebox{.98\linewidth}{!}{
  \begin{threeparttable}
  \begin{tabular}{cccccc}
    \toprule
    \toprule
    % & \multicolumn{3}{c}{Traffic congestion (5 nodes)}\\
    % \cmidrule(lr){2-4} \cmidrule(lr){5-7}  \cmidrule(lr){8-10} \cmidrule(lr) {11-13}
    Model & MAE (rare) ($\downarrow$) & MAE (frequent) ($\downarrow$) & MAE (total) ($\downarrow$) & Training log-likelihood ($\uparrow$) & AIC ($\downarrow$) \\
    \midrule
    % \texttt{Persistence Forecast} & 0.875 & 4.863 & 22.423 & / & /\\
    % \texttt{VAR} & & & & / \\
    % \midrule
    \texttt{Persistent} & 0.998 & 5.736 & 31.538 & / & / \\
    \texttt{VAR} & 0.906 & {\bf 3.680} & 21.940 & / & / \\
    \texttt{ETAS} & 0.785 & 4.266 & 30.925 & -2.476 & 45039.270\\
    \texttt{STNPP-GAT} & 0.728 & 3.875 & 21.561 & -2.427 & 44173.386\\
    \texttt{STNPP} & {\bf 0.716} & 3.708 & {\bf 20.080} & {\bf -2.413} & {\bf 44099.266}\\
    \bottomrule
    \bottomrule
  \end{tabular}
  % \begin{tablenotes}
  % \item *Numbers in parentheses are standard errors for three independent runs.
  % \end{tablenotes}
  \end{threeparttable}
  }
  % \vspace{-0.2in}
  \label{tab:in-sample-results}
\end{table}

More quantitative results about the in-sample estimations are summarized in Table~\ref{tab:in-sample-results}. To showcase our model's versatility in handling different types of events with distinct underlying mechanisms, we present separate in-sample estimation MAE assessments for events characterized by frequent or rare occurrences of the marks. The frequent event marks include those with landmark categories of ``financial,'' ``industrial,'' and ``restaurant,'' corresponding to the crime-landmark categories with the top nine total observations. The remaining crime-landmark categories are treated as rare event marks. We report MAE (rare), MAE (frequent), and MAE (total) as the final metrics, representing the estimation MAEs averaged over rare, frequent, and all event marks. The results in Table~\ref{tab:in-sample-results} demonstrate the comparable or superior predictive performance of \texttt{STNPP} against baselines. Note that although \texttt{VAR} has performance metrics that are close to our method, it is a time series model for predicting the event numbers, and it is not designed for dealing with discrete spatio-temporal events or providing any insights on the underlying event dynamics.

Table~\ref{tab:in-sample-results} also reports the training log-likelihood and the model AIC for three spatio-temporal point processes (we omit the comparison with AIC of \texttt{VAR}, which is not meaningful). The highest training log-likelihood and the lowest model AIC show that \texttt{STNPP} enjoys the best goodness-of-fit to the data.
Besides, the improved performance from \texttt{ETAS} to \texttt{STNPP-GAT} highlights the advantages of using street-network distance, and the performance gain of \texttt{STNPP} against \texttt{STNPP-GAT} emphasizes the benefits of incorporating nodal (mark) features in capturing complex event dynamics.

% We not only care about the estimation of the total number of events but also focus on the model's goodness-of-fit on events with different types of marks

\subsection{Out-of-sample predictions on testing data}

\begin{table}[!t]
  \caption{Quantitative results of out-of-sample estimation. Bold indicates the best performance.}
  % \vspace{-0.1in}
  \centering
  \resizebox{.98\linewidth}{!}{
  \begin{threeparttable}
  \begin{tabular}{ccccc}
    \toprule
    \toprule
    % & \multicolumn{3}{c}{Traffic congestion (5 nodes)}\\
    % \cmidrule(lr){2-4} \cmidrule(lr){5-7}  \cmidrule(lr){8-10} \cmidrule(lr) {11-13}
    Model & MAE (rare) ($\downarrow$) & MAE (frequent) ($\downarrow$) & MAE (total) ($\downarrow$) & Testing log-likelihood ($\uparrow$) \\
    \midrule
    % \texttt{Persistence Forecast} & 1.006 & 5.803 & 28.807 & / \\
    % \texttt{VAR} & & & & / \\
    % \midrule
    \texttt{Persistent} & 1.006 & 5.803 & 28.808 & / \\
    \texttt{VAR} & 0.998 & 5.502 & 27.507 & / \\
    \texttt{ETAS} & 0.879 & 4.786 & 28.715 & -2.223 \\
    \texttt{STNPP-GAT} & {\bf 0.769} & 4.329 & 26.302 & -2.201\\
    \texttt{STNPP} & 0.773 & {\bf 4.223} & {\bf 21.788} & {\bf -2.183} \\
    \bottomrule
    \bottomrule
  \end{tabular}
  % \begin{tablenotes}
  % \item *Numbers in parentheses are standard errors for three independent runs.
  % \end{tablenotes}
  \end{threeparttable}
  }
  % \vspace{-0.2in}
  \label{tab:out-of-sample-results}
\end{table}

\begin{figure}[!t]
%\vspace{-.15in}
\centering
\begin{subfigure}[h]{.32\linewidth}
\includegraphics[width=\linewidth]{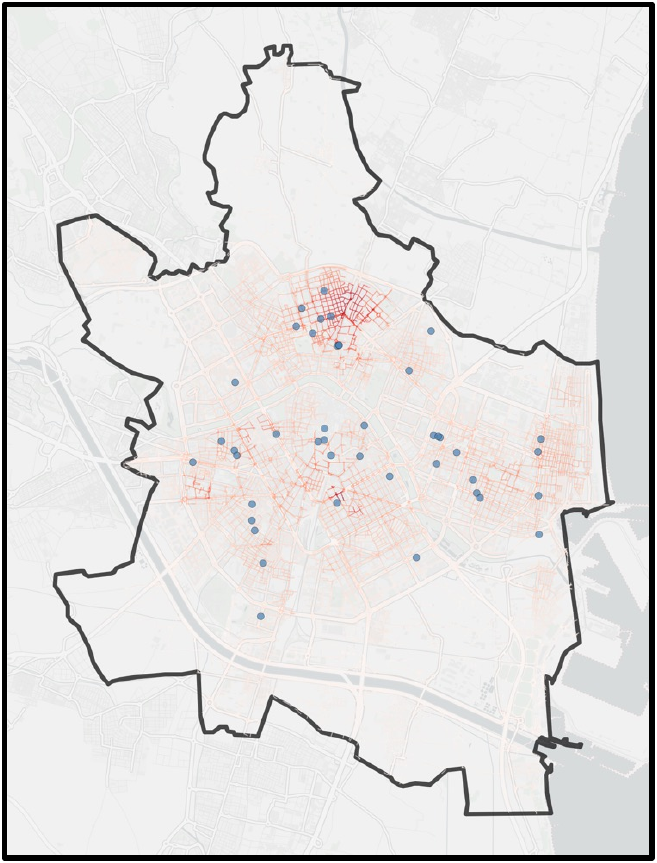}
\caption{\textit{Assault}, Apr. 11, 2019}
\end{subfigure}
\begin{subfigure}[h]{.32\linewidth}
\includegraphics[width=\linewidth]{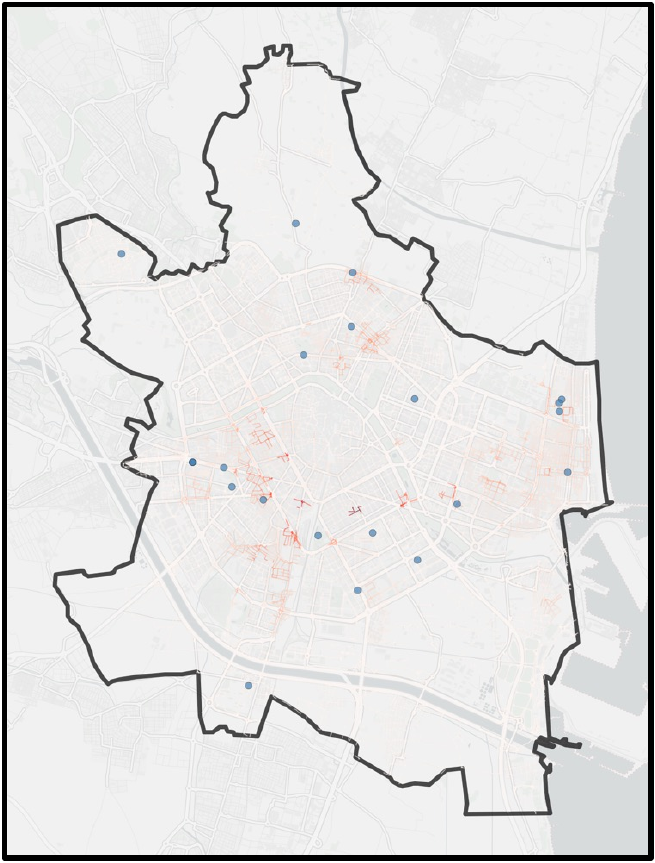}
\caption{\textit{Others}, July 20, 2019}
\end{subfigure}
\begin{subfigure}[h]{.32\linewidth}
\includegraphics[width=\linewidth]{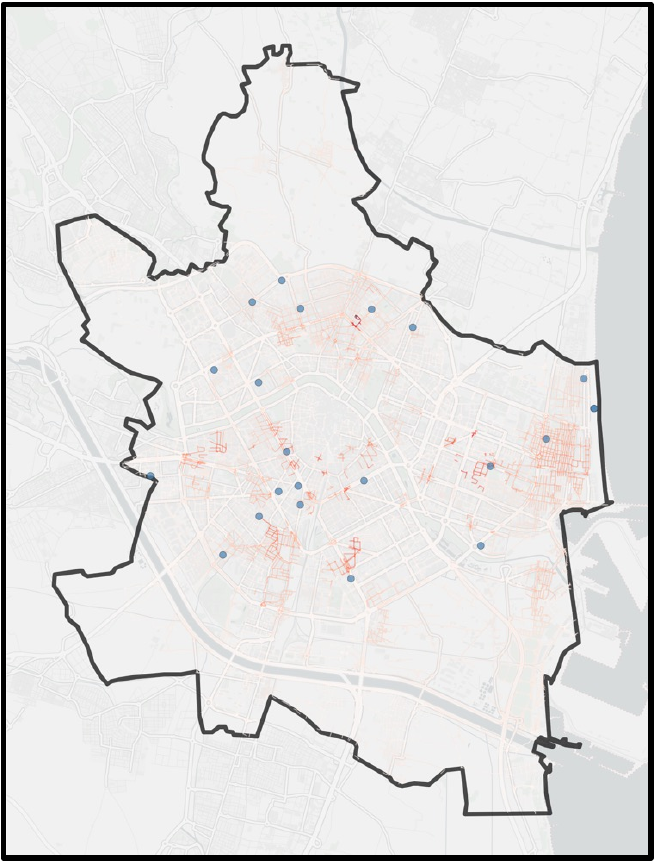}
\caption{\textit{Subtraction}, Oct. 28, 2019}
\end{subfigure}
\caption{Three snapshots of the out-of-sample prediction for the event intensity over the street network by \texttt{STNPP}. Each panel shows the predicted intensity of one type of crime on a given date in 2019. The depth of the red color indicates the value of the conditional intensity, and a deeper red color means a higher likelihood of future event occurrence. The blue dots represent actual incidents reported in the next two days from the given date. Our model provides intensity predictions that align well with the true observations.
}
\label{fig:out-of-sample-predicted-intensity}
\vspace{-0.1in}
\end{figure}

The model's predictive power can be assessed by the out-of-sample prediction task on the testing data set. We perform a one-week-ahead prediction of the number of events over the time window of 2019. Specifically, at a given time $t^*$ in 2019, we feed the data before $t^*$ into the learned model and evaluate the conditional intensity function over the next week. The predicted number of events with mark $c \times l$ in the following week can be estimated by the integral of the evaluated intensity function $\hat{\lambda}_{cl}$ over time and space, similarly as those in the in-sample predictions. The MAE between the predicted event numbers and the number of true observations are computed to indicate the model's predictive performance. We perform the out-of-sample prediction on a weekly basis over the year of 2019 and report the average prediction MAE. Table~\ref{tab:out-of-sample-results} presents the average MAEs of the predictions for the number of rare events, frequent events, and total events by our model \texttt{STNPP} and four baselines, indicating the superior performance of our model against other baselines on predicting the future. Besides the MAE, we also compare the fitted log-likelihood of the testing data using different point process models and report them in the table. The highest log-likelihood of \texttt{STNPP} showcases the best generalization ability of our model to the unseen data.

We visualize the predictive power of \texttt{STNPP} in Fig~\ref{fig:out-of-sample-predicted-intensity} by showing the predicted conditional intensity function for three types of crimes over the street network at different times in 2019. Each panel compares the predicted conditional intensity of one type of crime over space given the observed history with the true distribution of that type of crime in the next two days. As we can observe, the predicted event intensity by our model is consistent with the true distribution of future events, showing a higher intensity in those areas with a higher likelihood of observing crime events. Meanwhile, our model discerns the spatial patterns of different crimes by learning from the historical data, such as the risk for \textit{Assault} victims in major busy areas (\textit{e.g.}, the financial zone in the north part of the city) and a more regional, concentrated pattern for \textit{Subtraction} and \textit{Others}.

\subsection{Learned coefficients of mark interactions}

% Thresholding for the coefficients \citep{yuan2019multivariate}.

\begin{figure}[!t]
%\vspace{-.15in}
\centering
% \begin{subfigure}[h]{.48\linewidth}
% \includegraphics[width=\linewidth]{STNPP_alpha_matrix.pdf}
% \caption{Coefficients (\texttt{STNPP})}
% \end{subfigure}
% \begin{subfigure}[h]{.48\linewidth}
% \includegraphics[width=\linewidth]{STNPP-GAT_alpha_matrix.pdf}
% \caption{Coefficients (\texttt{STNPP-GAT})}
% \end{subfigure}
\begin{subfigure}[h]{.47\linewidth}
\includegraphics[width=\linewidth]{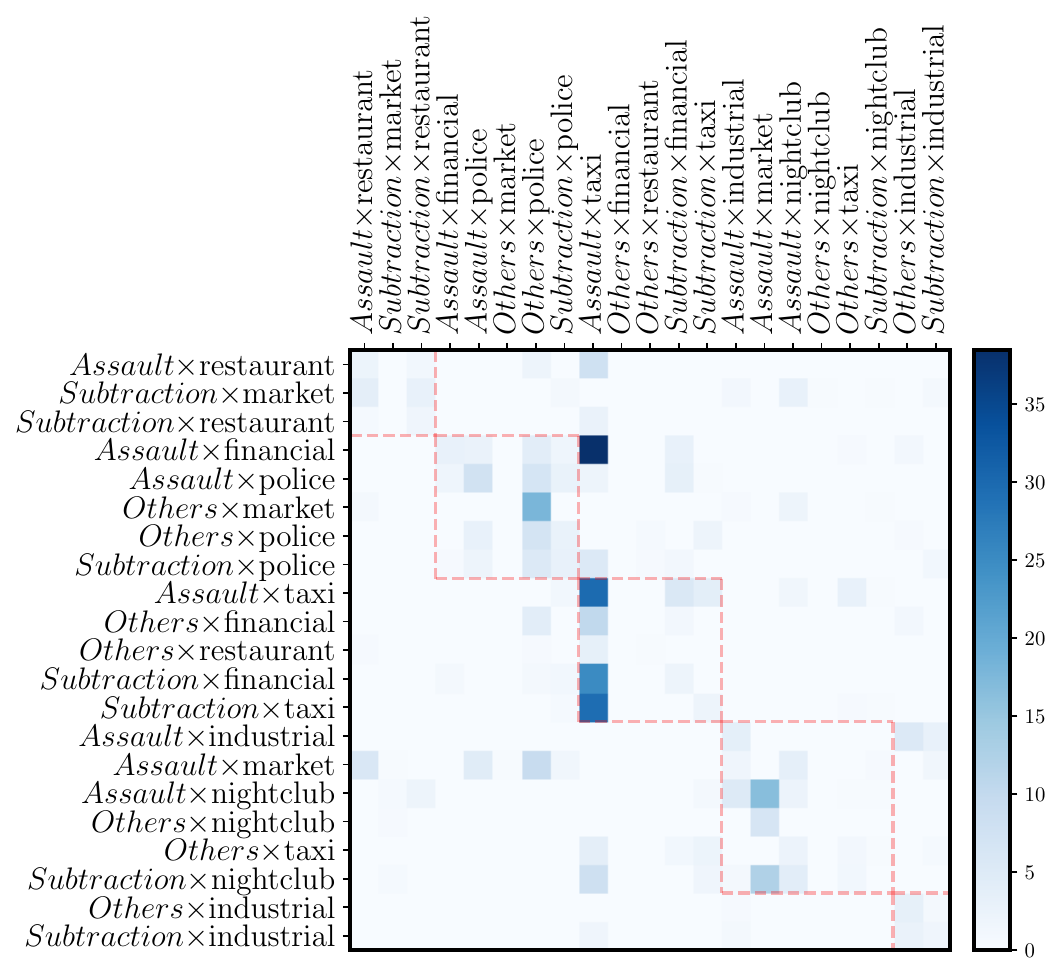}
\caption{Coefficients (\texttt{STNPP})}
\end{subfigure}
\begin{subfigure}[h]{.47\linewidth}
\includegraphics[width=\linewidth]{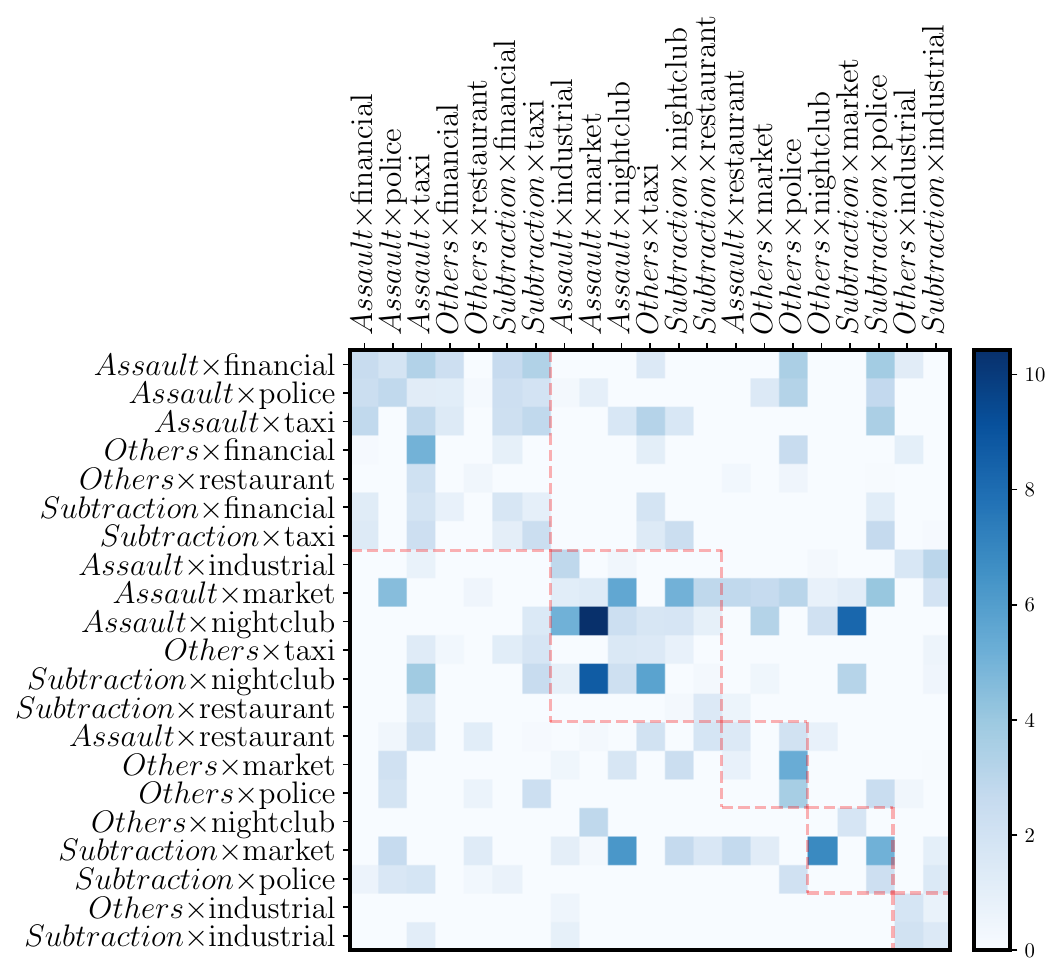}
\caption{Coefficients (\texttt{STNPP-GAT})}
\end{subfigure}

\begin{subfigure}[h]{.47\linewidth}
\includegraphics[width=\linewidth]{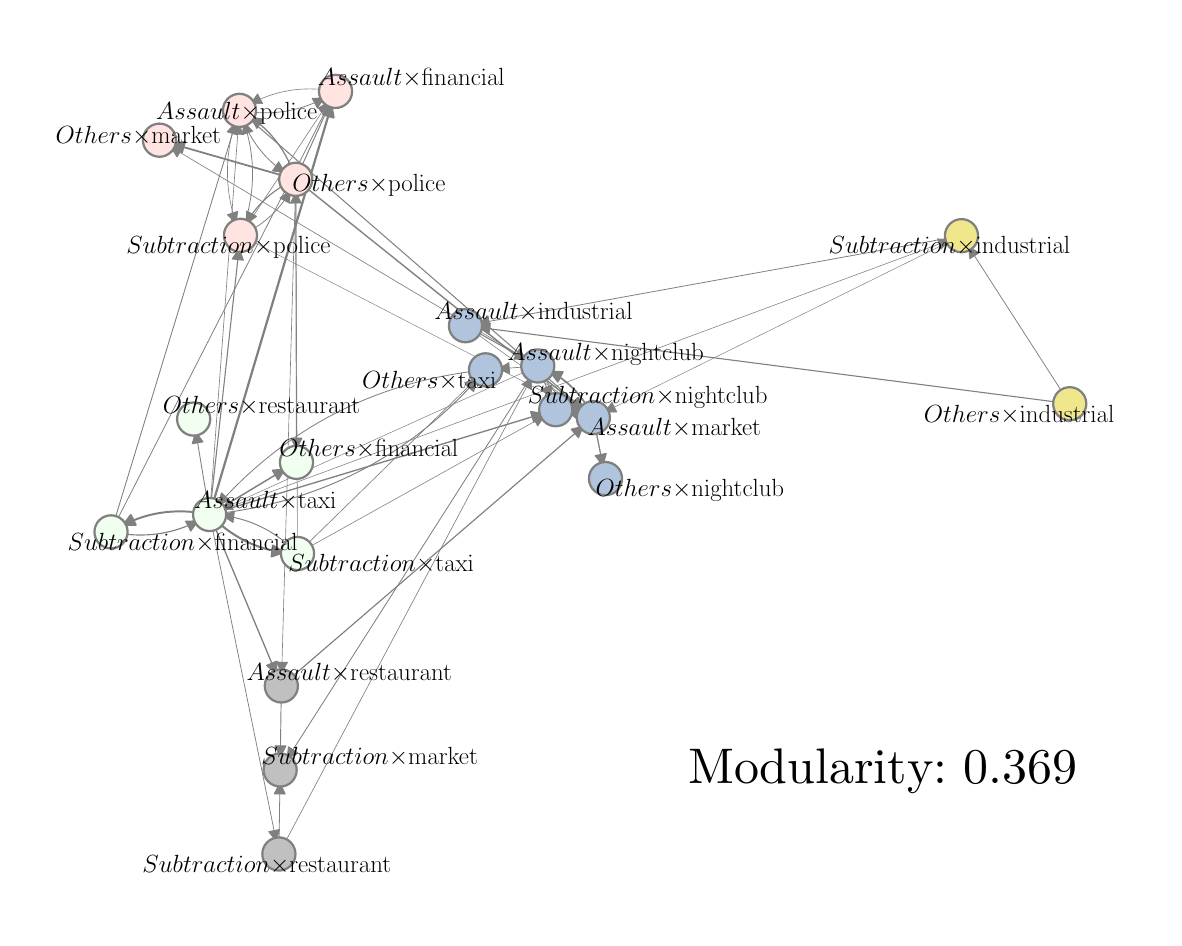}
\caption{Mark network (\texttt{STNPP}).}
\end{subfigure}
\begin{subfigure}[h]{.47\linewidth}
\includegraphics[width=\linewidth]{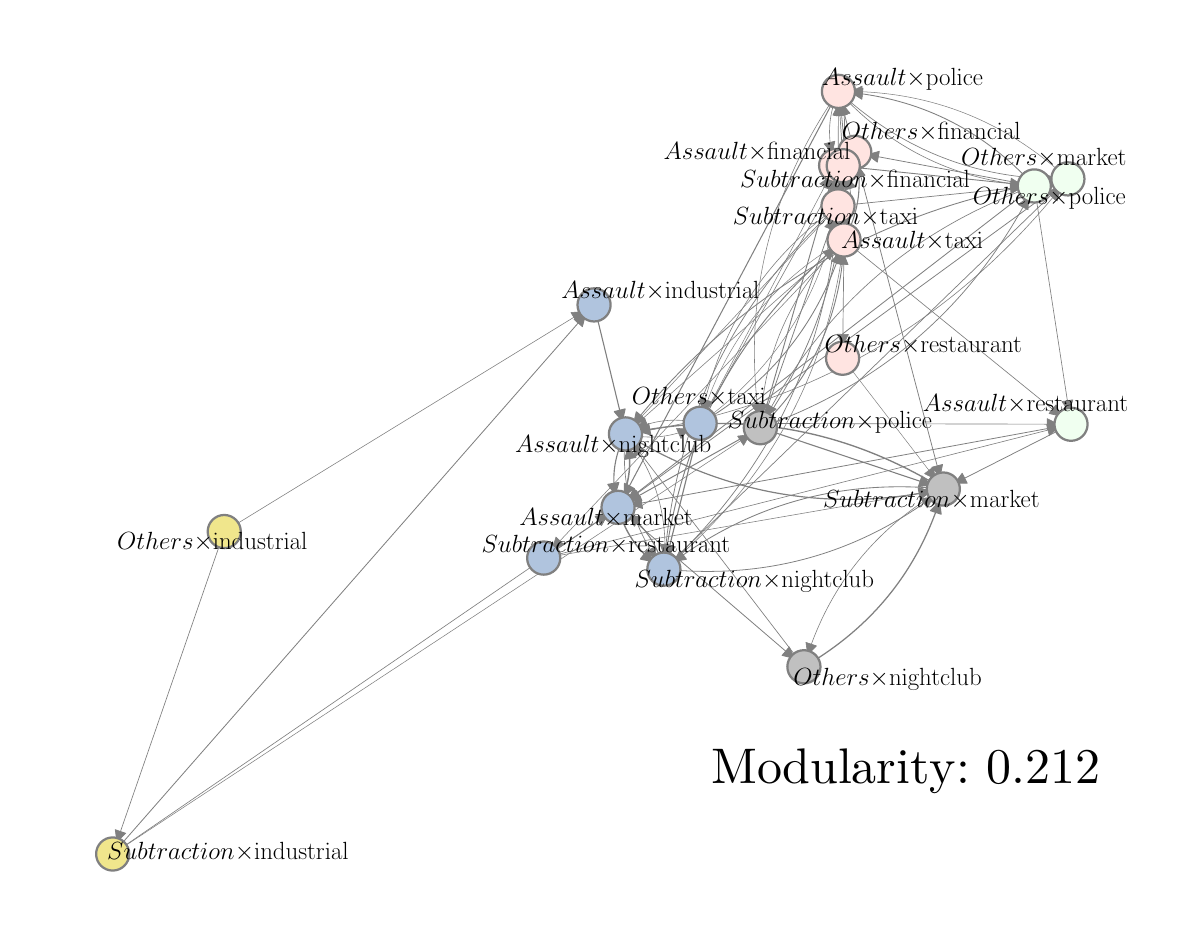}
\caption{Mark network (\texttt{STNPP-GAT}).}
\end{subfigure}
% \vspace{0.1in}
% \begin{subfigure}[h]{.95\linewidth}
% \includegraphics[width=\linewidth]{STNPP_alpha_matrix_trigger_event_num.pdf}
% \caption{Expected number of future events triggered by one event (\texttt{STNPP})}
% \end{subfigure}
\vspace{-5pt}
\caption{Learned mark interactions. (a)(b) Coefficients learned by \texttt{STNPP} and \texttt{STNPP-GAT}, respectively. The red dashed lines indicate communities detected by the Louvain algorithm based on the coefficients.
(c)(d) Mark networks learned by \texttt{STNPP} and \texttt{STNPP-GAT}. Nodes represent marks (one color means one community), and edges represent the interactions among marks (the arrow and line width indicate the direction and magnitude of the interaction).
% (e) Expected number of events triggered by one observed event with different marks.
}
\label{fig:alpha-matrix}
\vspace{-0.1in}
\end{figure}

\begin{figure}[!t]
%\vspace{-.15in}
\centering
\begin{subfigure}[h]{.95\linewidth}
\includegraphics[width=\linewidth]{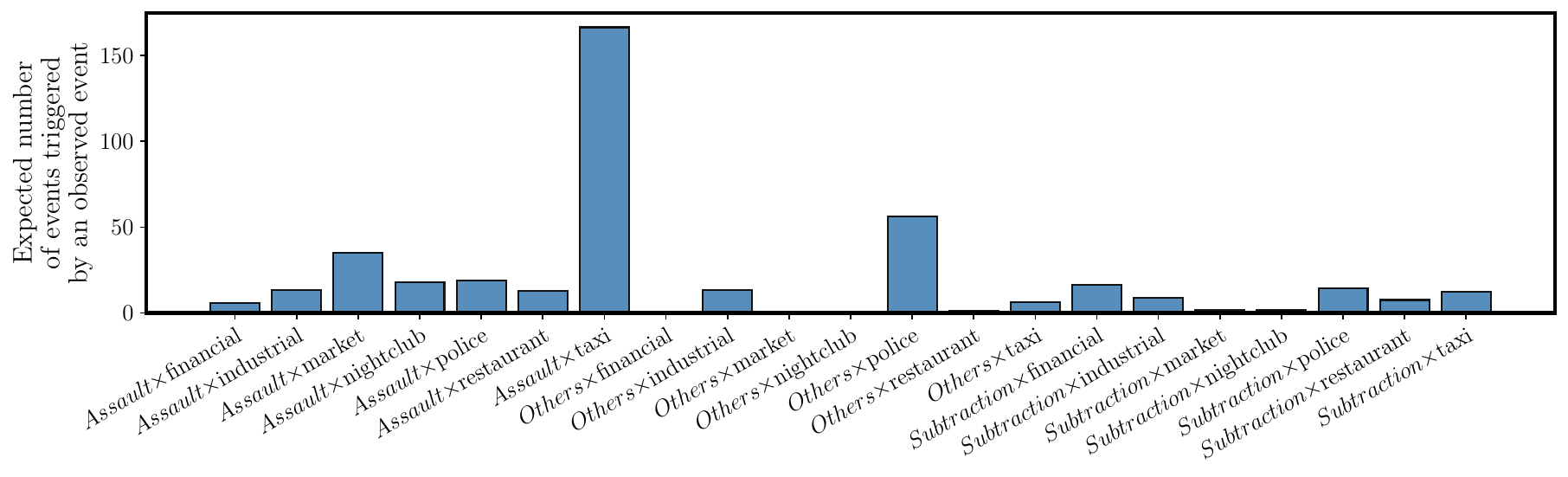}
\end{subfigure}
\caption{Expected number of events triggered by one observed event with different marks.
}
\label{fig:alpha-matrix-column-sum}
\vspace{-0.1in}
\end{figure}

The coefficients $\{\alpha_{cl, c'l'}\}_{c,c'\in\mathscr{C}, l,l'\in \mathscr{L}}$ learned by GAT capture the direction and magnitude of the influence between different event marks and is crucial in interpreting the model in practice. We visualize the learned coefficients by our model \texttt{STNPP} in Fig~\ref{fig:alpha-matrix}(a) by stacking them together into a matrix. 
Each matrix entry represents the coefficient that models the triggering effect from the event mark at the corresponding column to the mark at the corresponding row.
% The row and column of the matrix indexed by the first and second index of the coefficient. 
% When visualizing, we threshold the coefficients and only show the non-zero entries if $\alpha_{cl, c'l'} > 0.1$.
As the matrix can be regarded as the weighted adjacency matrix of the mark network we established in Section~\ref{sec:multiple-network-space}, we adopt the Louvain algorithm \citep{blondel2008fast} to perform community detection on the event marks. The detected communities tell us the groups of marks that are more closely connected, which are indicated by the square frames with red dashed lines in the visualized matrix.
% and those marks within the same community exhibit denser interactions among them than those with marks in other communities.
Five communities are detected based on the coefficients, suggesting different types of human daily activities. For instance, the largest community with six marks, including \textit{Assault} and \textit{Subtraction} in industrial, market, and nightclub zones, showcases the clustering patterns of certain crime events related to citizen activities after hours, such as grocery shopping or night amusement. Other communities also reveal criminal activities that are relevant to specific urban facilities, such as restaurants (the first community) and industrial zones (the last community).

We also visualize the mark network $A$ established from the learned coefficients in Figure~\ref{fig:alpha-matrix}(c), with nodes representing the event marks and edges indicating their interactions. The colors of the nodes suggest the detected communities of different marks. 
To demonstrate the benefits of adopting GAT to learn the coefficient and their community structure, we compare the learned coefficients by the ablated model \texttt{STNPP-GAT} in Fig~\ref{fig:alpha-matrix}(b)(d) with detected communities.
Although we have no ground truth to validate the community detection results, we here report the modularity \citep{newman2010networks} of the mark networks. Networks with higher modularity have stronger intra-community connections and fewer inter-community connections.
The modularity of the learned mark network by \texttt{STNPP} is much higher than the one learned by \texttt{STNPP-GAT}, as reported in Fig~\ref{fig:alpha-matrix}(c)(d). These visualizations also reveal a more distinct community structure in the network learned by \texttt{STNPP}, in contrast to the one of \texttt{STNPP-GAT}, which has more blurred community divisions. This high modularity of the mark network is beneficial for the decision-making of the local police department. For example, the detected communities highlight those closely connected marks and help identify influential crime events within specific communities. These insights can lead to more targeted and effective police patrolling against criminal activities.

To identify the most influential event marks, we plot the expected number of events that are triggered by an observed event with each mark in Fig~\ref{fig:alpha-matrix-column-sum}. The number of triggered events by one event with mark $c'\times l'$ is calculated by aggregating the coefficient $\alpha_{cl, c'l'}$ over the index $c$ and $l$, \textit{i.e.}, $\sum_{c \in \mathscr{C}, l \in \mathscr{L}}\alpha_{cl, c'l'}$, and a larger number indicates a stronger influence by an observed event with mark $c'\times l'$.
As we can see, crime events with marks of \textit{Assault}$\times$taxi have the strongest influence on the future by triggering the most number of events. Although this event mark is barely observed during the five-year period, its impact on subsequent event occurrences is not negligible.
% A reduced version of the coefficients by removing these three dominant marks and applying the Louvain algorithm \citep{blondel2008fast} for community detection on the remaining marks is shown in Figure~\ref{fig:alpha-matrix}(b). 
Other influential event marks include those related to police zones or assault activities, indicating the heterogeneity of the event dynamics across different crime types and urban areas.

\begin{figure}[!t]
%\vspace{-.15in}
\centering
\begin{subfigure}[h]{.95\linewidth}
\includegraphics[width=\linewidth]{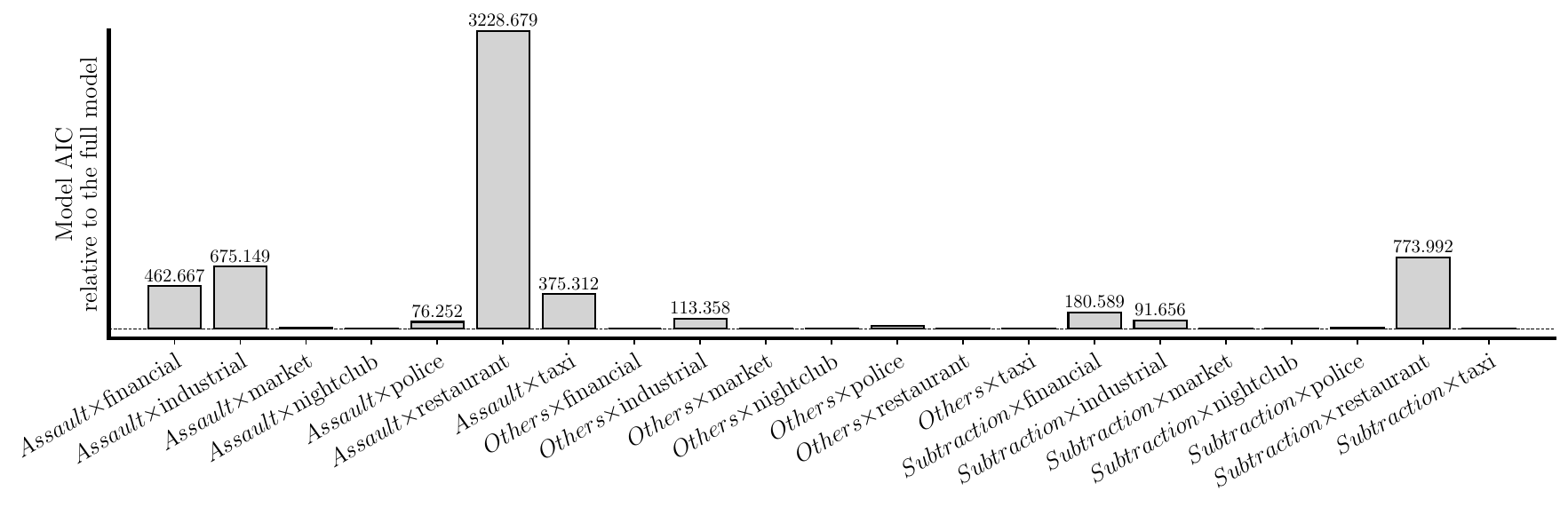}
\caption{Degradation in AIC of each reduced model.}
\end{subfigure}

\begin{subfigure}[h]{.95\linewidth}
\includegraphics[width=\linewidth]{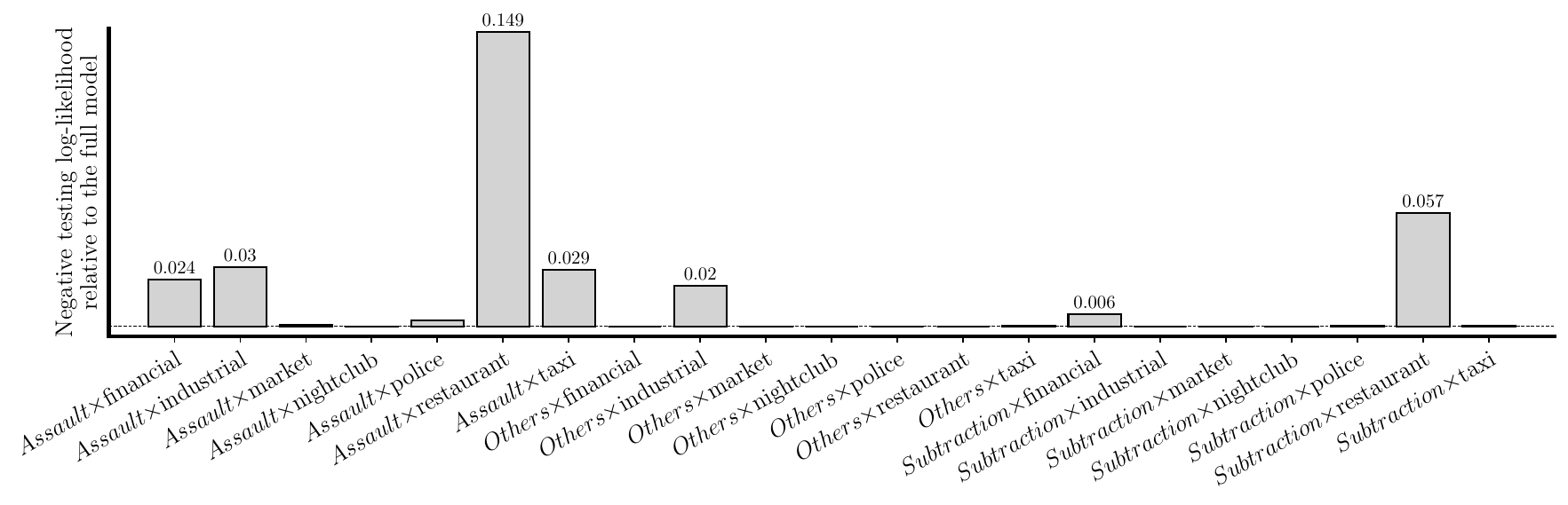}
\caption{Degradation in negative testing log-likelihood of each reduced model.}
\end{subfigure}

\begin{subfigure}[h]{.95\linewidth}
\includegraphics[width=\linewidth]{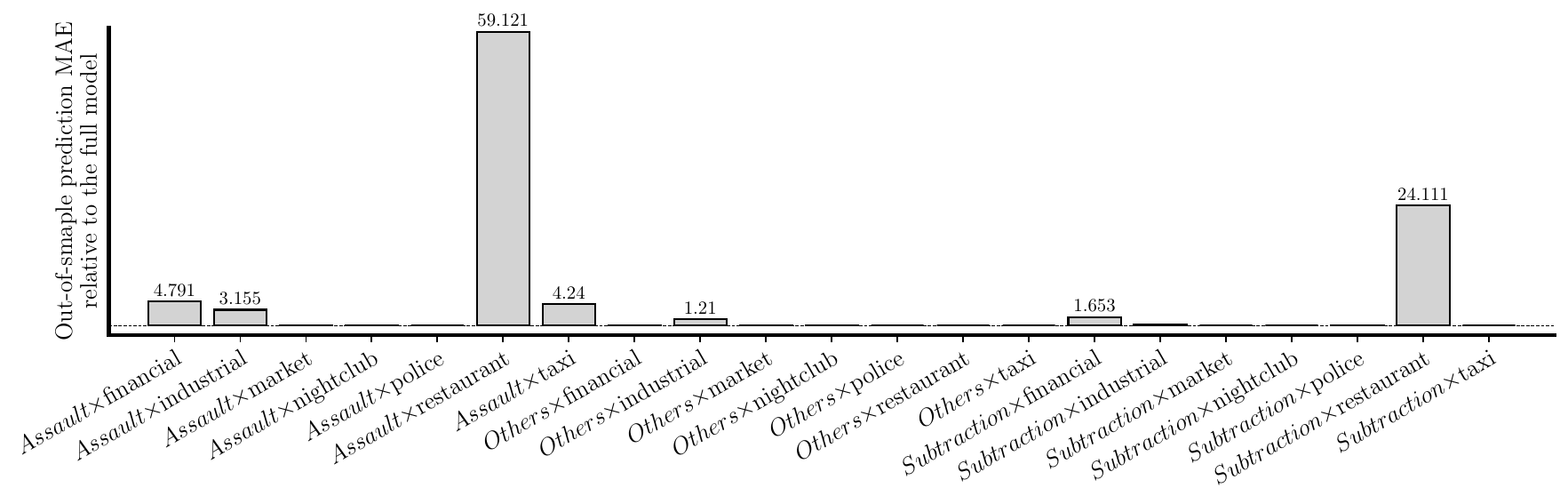}
\caption{Degradation in out-of-sample prediction MAE of each reduced model.}
\end{subfigure}
\caption{Performance degradation of different reduced models by zeroing out the influence of events with different crime-landmark labels. The horizontal axis indicates the type of events whose influence is removed from the full model. For each reduced model, we evaluate the testing log-likelihood and out-of-sample prediction MAE on the data in 2019. The height of the bars represents the difference between the metric scores of the reduced model and the full model. The dashed line in each panel indicates zero value. A larger value of difference indicates a higher level of importance of the crime-landmark label in the full model.
}
\label{fig:reduced-models}
\vspace{-0.1in}
\end{figure}

\subsection{Important event marks}

Another important task for local practitioners in implementing effective prevention strategies is to identify particular types of crime events that can lead to an obvious risk increase in the community's exposure to the crimes. 
These events not only include those that trigger the subsequent event occurrences to a large extent, such as \textit{Assault}$\times$taxi, but also include those event types that have a smaller influence magnitude on future events but are frequently reported, such as \textit{Assault}$\times$restaurant. 
% Therefore, analysis involving both the influence and the frequency of different types of events is imperative to accurately identify those important marks in the underlying event generation mechanisms.

To this end, we investigate the contribution of each event mark $c' \times l'$ to the underlying event generation mechanisms by neutralizing its influence (\textit{i.e.}, set $\alpha_{cl, c'l'} = 0, \forall c\in \mathscr{C}, l \in \mathscr{L}$) and then evaluating the performance gap between the reduced model and the original model. 
A larger performance gap indicates a higher importance of that event mark in the effectiveness of the model. We evaluate the performance of each reduced model with influence from one type of event mark neutralized (a total of 21 reduced models) in terms of three metrics: AIC on training data, negative log-likelihood on testing data, and out-of-sample prediction MAE, and compare them with the metrics obtained by the full model. The corresponding results are presented in Fig~\ref{fig:reduced-models}. 
For the top two marks that have the most expected number of triggered events shown in Fig~\ref{fig:alpha-matrix-column-sum}, the neutralization of the influence of \textit{Assault}$\times$taxi leads to obvious performance degradation, while the other one of \textit{Others}$\times$police have a much smaller impact, due to the scarce observation of this event mark during the investigation period.
Other event marks, to which the neutralization of the influence can significantly decline the model performance, include those of \textit{Assault}$\times$financial, \textit{Assault}$\times$industrial, \textit{Assault}$\times$restaurant, \textit{Subtraction}$\times$restaurant, and so on. Crime events with these marks relate to those places and the daily activities of citizens who are more vulnerable to criminals. These events are more frequently observed than others, and their cumulative effects on attracting future criminal activities need to be paid attention to.

\section{Discussion}

We have presented a new spatio-temporal-network point process developed to model crime events within Valencia, Spain. This model is built on the city’s street network for spatial analysis, mirroring the real-world context where urban crimes mainly occur along streets. The introduction of a spatial kernel that measures distance across the street network respects the intrinsic network nature of urban areas, capturing the contagion effect of crime more accurately and realistically compared to traditional point process models with kernels that rely on Euclidean distance.
The integration of urban environmental factors such as nearby facilities and land use into our analysis adds another layer of depth. By partitioning the city into different functional zones and creating new event marks with corresponding crime and zone categories, our model allows us to explore how specific urban environments foster particular types of crime. The adoption of a graph attention neural network architecture improves the learning of the complex interactions between various event marks, which also enables the identification of those important crime types in different environmental contexts, leading to insights that could inform targeted interventions.
The numerical results on the real crime data in Valencia demonstrate the superior performance of our model against common baselines in forecasting the numbers of crime events and their distributions. The results not only prove the effectiveness of our model in actual practice but also underscore the importance of models tailored to crime modeling in specific urban contexts.

Several avenues exist for further enhancement of our model. Considering a directed street network could offer additional insights, particularly in scenarios where the movement direction of perpetrators (such as those in vehicles) plays a role in crime execution. A more rigorous statistical analysis of the significance of learned mark-to-mark interactions would enhance the robustness of our findings, potentially revealing more intricate patterns that could be pivotal for law enforcement and urban planning strategies.
Another future direction is to conduct a systematic analysis of spatial covariate effects to explain variations in crime intensity and clustering when such data are available. Extending our spatio-temporal-network framework to jointly model these covariates could further enhance its explanatory power and policy relevance.
By addressing these areas, we aim to refine our understanding of urban crime dynamics further, thus not only contributing to academic discourse but also providing a practical framework for enhancing public safety and security in urban settings.

\section*{Acknowledgement}

This work is partially supported by an NSF CAREER CCF-1650913, NSF DMS-2134037, CMMI-2015787, CMMI-2112533, DMS-GR00023160, DMS-1938106, DMS-1830210, ONR N000142412278, and the Coca-Cola Foundation.

% \jm{NEED TO BE ENLARGED}

%%%%%%%%%%%%%%%%%%%%%%%%%%%%%%%%%%%%%%%%%%%%%%
%% Support information, if any,             %%
%% should be provided in the                %%
%% Acknowledgements section.                %%
%%%%%%%%%%%%%%%%%%%%%%%%%%%%%%%%%%%%%%%%%%%%%%
%\begin{acks}[Acknowledgments]
% The authors would like to thank ...
%\end{acks}
%%%%%%%%%%%%%%%%%%%%%%%%%%%%%%%%%%%%%%%%%%%%%%
%% Funding information, if any,             %%
%% should be provided in the                %%
%% funding section.                         %%
%%%%%%%%%%%%%%%%%%%%%%%%%%%%%%%%%%%%%%%%%%%%%%
%\begin{funding}
% The first author was supported by ...
%
% The second author was supported in part by ...
%\end{funding}

%%%%%%%%%%%%%%%%%%%%%%%%%%%%%%%%%%%%%%%%%%%%%%
%% Supplementary Material, including data   %%
%% sets and code, should be provided in     %%
%% {supplement} environment with title      %%
%% and short description. It cannot be      %%
%% available exclusively as external link.  %%
%% All Supplementary Material must be       %%
%% available to the reader on Project       %%
%% Euclid with the published article.       %%
%%%%%%%%%%%%%%%%%%%%%%%%%%%%%%%%%%%%%%%%%%%%%%
%\begin{supplement}
%\stitle{???}
%\sdescription{???.}
%\end{supplement}

\bibliographystyle{apalike}
\bibliography{arxiv_refs}

\begin{thebibliography}{}

\bibitem[Akaike, 1974]{akaike1974new}
Akaike, H. (1974).
\newblock A new look at the statistical model identification.
\newblock {\em IEEE transactions on automatic control}, 19(6):716--723.

\bibitem[Akaike, 1998]{akaike1998information}
Akaike, H. (1998).
\newblock Information theory and an extension of the maximum likelihood
  principle.
\newblock In {\em Selected papers of hirotugu akaike}, pages 199--213.
  Springer.

\bibitem[Andresen, 2007]{andresen2007location}
Andresen, M.~A. (2007).
\newblock Location quotients, ambient populations, and the spatial analysis of
  crime in vancouver, canada.
\newblock {\em Environment and Planning A}, 39(10):2423--2444.

\bibitem[Blondel et~al., 2008]{blondel2008fast}
Blondel, V.~D., Guillaume, J.-L., Lambiotte, R., and Lefebvre, E. (2008).
\newblock Fast unfolding of communities in large networks.
\newblock {\em Journal of statistical mechanics: theory and experiment},
  2008(10):P10008.

\bibitem[Bonta et~al., 1998]{bonta1998prediction}
Bonta, J., Law, M., and Hanson, K. (1998).
\newblock The prediction of criminal and violent recidivism among mentally
  disordered offenders: a meta-analysis.
\newblock {\em Psychological bulletin}, 123(2):123.

\bibitem[Bounce, 2024]{bounceValencia}
Bounce (2024).
\newblock Is valencia safe to visit? a comprehensive safety guide.

\bibitem[Bowers et~al., 2004]{bowers2004prospective}
Bowers, K.~J., Johnson, S.~D., and Pease, K. (2004).
\newblock Prospective hot-spotting: the future of crime mapping?
\newblock {\em British journal of criminology}, 44(5):641--658.

\bibitem[Browning et~al., 2010]{browning2010commercial}
Browning, C.~R., Byron, R.~A., Calder, C.~A., Krivo, L.~J., Kwan, M.-P., Lee,
  J.-Y., and Peterson, R.~D. (2010).
\newblock Commercial density, residential concentration, and crime: Land use
  patterns and violence in neighborhood context.
\newblock {\em Journal of Research in Crime and Delinquency}, 47(3):329--357.

\bibitem[Cai et~al., 2024]{cai2024latent}
Cai, B., Zhang, J., and Guan, Y. (2024).
\newblock Latent network structure learning from high-dimensional multivariate
  point processes.
\newblock {\em Journal of the American Statistical Association},
  119(545):95--108.

\bibitem[Chainey et~al., 2008]{chainey2008utility}
Chainey, S., Tompson, L., and Uhlig, S. (2008).
\newblock The utility of hotspot mapping for predicting spatial patterns of
  crime.
\newblock {\em Security journal}, 21:4--28.

\bibitem[Cheng et~al., 2025]{cheng2025deep}
Cheng, X., Dong, Z., and Xie, Y. (2025).
\newblock Deep spatio-temporal point processes: Advances and new directions.
\newblock {\em arXiv preprint arXiv:2504.06364}.

\bibitem[Cho et~al., 2013]{cho2013latent}
Cho, Y.-S., Galstyan, A., Brantingham, P.~J., and Tita, G. (2013).
\newblock Latent self-exciting point process model for spatial-temporal
  networks.
\newblock {\em arXiv preprint arXiv:1302.2671}.

\bibitem[Daley et~al., 2003]{daley2003introduction}
Daley, D.~J., Vere-Jones, D., et~al. (2003).
\newblock {\em An introduction to the theory of point processes: volume I:
  elementary theory and methods}.
\newblock Springer.

\bibitem[Dong et~al., 2023a]{dong2023spatiotemporal}
Dong, Z., Cheng, X., and Xie, Y. (2023a).
\newblock Spatio-temporal point processes with deep non-stationary kernels.
\newblock In {\em The Eleventh International Conference on Learning
  Representations}.

\bibitem[Dong et~al., 2023b]{dong2023deep}
Dong, Z., Repasky, M., Cheng, X., and Xie, Y. (2023b).
\newblock Deep graph kernel point processes.
\newblock In {\em Temporal Graph Learning Workshop @ NeurIPS 2023}.

\bibitem[Dong and Xie, 2024]{dong2024atlanta}
Dong, Z. and Xie, Y. (2024).
\newblock Atlanta gun violence modeling via nonstationary spatio-temporal point
  processes.
\newblock {\em arXiv preprint arXiv:2408.09258}.

\bibitem[Dong et~al., 2023c]{dong2023non}
Dong, Z., Zhu, S., Xie, Y., Mateu, J., and Rodr{\'\i}guez-Cort{\'e}s, F.~J.
  (2023c).
\newblock Non-stationary spatio-temporal point process modeling for
  high-resolution covid-19 data.
\newblock {\em Journal of the Royal Statistical Society Series C: Applied
  Statistics}, 72(2):368--386.

\bibitem[Du et~al., 2016]{du2016recurrent}
Du, N., Dai, H., Trivedi, R., Upadhyay, U., Gomez-Rodriguez, M., and Song, L.
  (2016).
\newblock Recurrent marked temporal point processes: Embedding event history to
  vector.
\newblock In {\em Proceedings of the 22nd ACM SIGKDD International Conference
  on Knowledge Discovery and Data mining}, pages 1555--1564.

\bibitem[D’Angelo et~al., 2024]{d2024self}
D’Angelo, N., Payares, D., Adelfio, G., and Mateu, J. (2024).
\newblock Self-exciting point process modelling of crimes on linear networks.
\newblock {\em Statistical Modelling}, 24(2):139--168.

\bibitem[Fang et~al., 2023]{fang2023group}
Fang, G., Xu, G., Xu, H., Zhu, X., and Guan, Y. (2023).
\newblock Group network hawkes process.
\newblock {\em Journal of the American Statistical Association}, pages 1--17.

\bibitem[Fleming et~al., 1994]{fleming1994exploring}
Fleming, Z., Brantingham, P., Brantingham, P., et~al. (1994).
\newblock Exploring auto theft in british columbia.
\newblock {\em Crime prevention studies}, 3:47--90.

\bibitem[Gao et~al., 2017]{gao2017extracting}
Gao, S., Janowicz, K., and Couclelis, H. (2017).
\newblock Extracting urban functional regions from points of interest and human
  activities on location-based social networks.
\newblock {\em Transactions in GIS}, 21(3):446--467.

\bibitem[Gottfredson, 1981]{gottfredson1981etiology}
Gottfredson, M.~R. (1981).
\newblock On the etiology of criminal victimization.
\newblock {\em J. Crim. L. \& Criminology}, 72:714.

\bibitem[Grann et~al., 1999]{grann1999psychopathy}
Grann, M., L{\aa}ngstr{\"o}m, N., Tengstr{\"o}m, A., and Kullgren, G. (1999).
\newblock Psychopathy ({PCL-R}) predicts violent recidivism among criminal
  offenders with personality disorders in sweden.
\newblock {\em Law and human behavior}, 23:205--217.

\bibitem[Groff, 2011]{groff2011exploring}
Groff, E. (2011).
\newblock Exploring ‘near’: Characterizing the spatial extent of drinking
  place influence on crime.
\newblock {\em Australian \& New Zealand Journal of Criminology},
  44(2):156--179.

\bibitem[Hawkes, 1971]{hawkes1971spectra}
Hawkes, A.~G. (1971).
\newblock Spectra of some self-exciting and mutually exciting point processes.
\newblock {\em Biometrika}, 58(1):83--90.

\bibitem[Hessellund et~al., 2022a]{hessellund2022second}
Hessellund, K.~B., Xu, G., Guan, Y., and Waagepetersen, R. (2022a).
\newblock Second-order semi-parametric inference for multivariate log gaussian
  cox processes.
\newblock {\em Journal of the Royal Statistical Society Series C: Applied
  Statistics}, 71(1):244--268.

\bibitem[Hessellund et~al., 2022b]{hessellund2022semiparametric}
Hessellund, K.~B., Xu, G., Guan, Y., and Waagepetersen, R. (2022b).
\newblock Semiparametric multinomial logistic regression for multivariate point
  pattern data.
\newblock {\em Journal of the American Statistical Association},
  117(539):1500--1515.

\bibitem[Hu and Han, 2019]{hu2019identification}
Hu, Y. and Han, Y. (2019).
\newblock Identification of urban functional areas based on poi data: A case
  study of the guangzhou economic and technological development zone.
\newblock {\em Sustainability}, 11(5):1385.

\bibitem[Jiang et~al., 2015]{jiang2015mining}
Jiang, S., Alves, A., Rodrigues, F., Ferreira~Jr, J., and Pereira, F.~C.
  (2015).
\newblock Mining point-of-interest data from social networks for urban land use
  classification and disaggregation.
\newblock {\em Computers, Environment and Urban Systems}, 53:36--46.

\bibitem[Johnson, 2008]{johnson2008repeat}
Johnson, S.~D. (2008).
\newblock Repeat burglary victimisation: a tale of two theories.
\newblock {\em Journal of Experimental Criminology}, 4:215--240.

\bibitem[Johnson and Bowers, 2010]{johnson2010permeability}
Johnson, S.~D. and Bowers, K.~J. (2010).
\newblock Permeability and burglary risk: Are cul-de-sacs safer?
\newblock {\em Journal of Quantitative Criminology}, 26:89--111.

\bibitem[Kennedy et~al., 2011]{kennedy2011risk}
Kennedy, L.~W., Caplan, J.~M., and Piza, E. (2011).
\newblock Risk clusters, hotspots, and spatial intelligence: risk terrain
  modeling as an algorithm for police resource allocation strategies.
\newblock {\em Journal of quantitative criminology}, 27:339--362.

\bibitem[Kennedy et~al., 2016]{kennedy2016vulnerability}
Kennedy, L.~W., Caplan, J.~M., Piza, E.~L., and Buccine-Schraeder, H. (2016).
\newblock Vulnerability and exposure to crime: Applying risk terrain modeling
  to the study of assault in chicago.
\newblock {\em Applied Spatial Analysis and Policy}, 9:529--548.

\bibitem[Kinney et~al., 2008]{kinney2008crime}
Kinney, J.~B., Brantingham, P.~L., Wuschke, K., Kirk, M.~G., and Brantingham,
  P.~J. (2008).
\newblock Crime attractors, generators and detractors: Land use and urban crime
  opportunities.
\newblock {\em Built environment}, 34(1):62--74.

\bibitem[Kivel{\"a} et~al., 2014]{kivela2014multilayer}
Kivel{\"a}, M., Arenas, A., Barthelemy, M., Gleeson, J.~P., Moreno, Y., and
  Porter, M.~A. (2014).
\newblock Multilayer networks.
\newblock {\em Journal of complex networks}, 2(3):203--271.

\bibitem[Kumazawa and Ogata, 2014]{kumazawa2014nonstationary}
Kumazawa, T. and Ogata, Y. (2014).
\newblock {Nonstationary ETAS models for nonstandard earthquakes}.
\newblock {\em The Annals of Applied Statistics}, 8(3):1825 -- 1852.

\bibitem[Lev-Wiesel et~al., 2004]{lev2004posttraumatic}
Lev-Wiesel, R., Amir, M., and Besser, A. (2004).
\newblock Posttraumatic growth among female survivors of childhood sexual abuse
  in relation to the perpetrator identity.
\newblock {\em Journal of Loss and Trauma}, 10(1):7--17.

\bibitem[Levine and CrimeStat, 2002]{levine2002spatial}
Levine, N. and CrimeStat, I. (2002).
\newblock A spatial statistics program for the analysis of crime incident
  locations.
\newblock {\em Ned Levine and Associates, Houston, TX, and the National
  Institute of Justice, Washington, DC}.

\bibitem[Lewis et~al., 2012]{lewis2012self}
Lewis, E., Mohler, G., Brantingham, P.~J., and Bertozzi, A.~L. (2012).
\newblock Self-exciting point process models of civilian deaths in {I}raq.
\newblock {\em Security Journal}, 25:244--264.

\bibitem[Li et~al., 2023]{li2023stochastic}
Li, J., Liu, X., Dahan, M., and Montreuil, B. (2023).
\newblock Stochastic service network design with different operational patterns
  for hyperconnected relay transportation.
\newblock In {\em Proceedings of 9th International Physical Internet Conference
  (IPIC)}.

\bibitem[Liao et~al., 2022]{liao2022tides}
Liao, C.-Y., Garcia, G.-G., Paynabar, K., Dong, Z., Xie, Y., and Jalali, M.~S.
  (2022).
\newblock Tides need stemmed: A locally operating spatio-temporal mutually
  exciting point process with dynamic network for improving opioid overdose
  death prediction.
\newblock {\em arXiv preprint arXiv:2211.07570}.

\bibitem[Linderman and Adams, 2014]{linderman2014discovering}
Linderman, S. and Adams, R. (2014).
\newblock Discovering latent network structure in point process data.
\newblock In {\em International Conference on Machine Learning}, pages
  1413--1421. PMLR.

\bibitem[Liu et~al., 2021]{liu2021point}
Liu, X., Carter, J., Ray, B., and Mohler, G. (2021).
\newblock {Point process modeling of drug overdoses with heterogeneous and
  missing data}.
\newblock {\em The Annals of Applied Statistics}, 15(1):88 -- 101.

\bibitem[Liu et~al., 2025a]{liu2025dynamic}
Liu, X., Li, J., Dahan, M., and Montreuil, B. (2025a).
\newblock Dynamic hub capacity planning in hyperconnected relay transportation
  networks under uncertainty.
\newblock {\em Transportation Research Part E: Logistics and Transportation
  Review}, 194:103940.

\bibitem[Liu et~al., 2025b]{liu2025network}
Liu, X., Muthukrishnan, P., and Montreuil, B. (2025b).
\newblock Network design and capacity management in hyperconnected urban
  logistic networks.
\newblock {\em Proceedings of 11th International Physical Internet Conference
  (IPIC)}.

\bibitem[Loeffler and Flaxman, 2018]{loeffler2018gun}
Loeffler, C. and Flaxman, S. (2018).
\newblock Is gun violence contagious? a spatiotemporal test.
\newblock {\em Journal of quantitative criminology}, 34:999--1017.

\bibitem[Long et~al., 2015]{long2015discovering}
Long, Y., Shen, Z., Long, Y., and Shen, Z. (2015).
\newblock Discovering functional zones using bus smart card data and points of
  interest in beijing.
\newblock {\em Geospatial analysis to support urban planning in Beijing}, pages
  193--217.

\bibitem[Maas et~al., 2013]{maas2013rectifier}
Maas, A.~L., Hannun, A.~Y., Ng, A.~Y., et~al. (2013).
\newblock Rectifier nonlinearities improve neural network acoustic models.
\newblock In {\em Proc. icml}, volume~30, page~3. Atlanta, GA.

\bibitem[Meera and Jayakumar, 1995]{meera1995determinants}
Meera, A.~K. and Jayakumar, M.~D. (1995).
\newblock Determinants of crime in a developing country: a regression model.
\newblock {\em Applied Economics}, 27(5):455--460.

\bibitem[Mei and Eisner, 2017]{mei2017neural}
Mei, H. and Eisner, J.~M. (2017).
\newblock The neural hawkes process: A neurally self-modulating multivariate
  point process.
\newblock {\em Advances in Neural Information Processing Systems}, 30.

\bibitem[Mohler, 2013]{mohler2013modeling}
Mohler, G. (2013).
\newblock Modeling and estimation of multi-source clustering in crime and
  security data.
\newblock {\em The Annals of Applied Statistics}, pages 1525--1539.

\bibitem[Mohler, 2014]{mohler2014marked}
Mohler, G. (2014).
\newblock Marked point process hotspot maps for homicide and gun crime
  prediction in chicago.
\newblock {\em International Journal of Forecasting}, 30(3):491--497.

\bibitem[Mohler et~al., 2011]{mohler2011self}
Mohler, G.~O., Short, M.~B., Brantingham, P.~J., Schoenberg, F.~P., and Tita,
  G.~E. (2011).
\newblock Self-exciting point process modeling of crime.
\newblock {\em Journal of the American Statistical Association},
  106(493):100--108.

\bibitem[Moller and Waagepetersen, 2003]{moller2003statistical}
Moller, J. and Waagepetersen, R.~P. (2003).
\newblock {\em Statistical inference and simulation for spatial point
  processes}.
\newblock CRC press.

\bibitem[Neill and Gorr, 2007]{neill2007detecting}
Neill, D.~B. and Gorr, W.~L. (2007).
\newblock Detecting and preventing emerging epidemics of crime.
\newblock {\em Advances in Disease Surveillance}, 4(13).

\bibitem[Newman, 2010]{newman2010networks}
Newman, M. (2010).
\newblock {\em {Networks: An Introduction}}.
\newblock Oxford University Press.

\bibitem[Ogata, 1988]{ogata1988statistical}
Ogata, Y. (1988).
\newblock Statistical models for earthquake occurrences and residual analysis
  for point processes.
\newblock {\em Journal of the American Statistical Association}, 83(401):9--27.

\bibitem[Ogata, 1998]{ogata1998space}
Ogata, Y. (1998).
\newblock Space-time point-process models for earthquake occurrences.
\newblock {\em Annals of the Institute of Statistical Mathematics},
  50:379--402.

\bibitem[{OpenStreetMap contributors}, 2017]{OpenStreetMap}
{OpenStreetMap contributors} (2017).
\newblock {Planet dump retrieved from https://planet.osm.org }.
\newblock \url{ https://www.openstreetmap.org }.

\bibitem[Perry, 2013]{perry2013predictive}
Perry, W.~L. (2013).
\newblock {\em Predictive policing: The role of crime forecasting in law
  enforcement operations}.
\newblock Rand Corporation.

\bibitem[Porter and White, 2012]{porter2012self}
Porter, M.~D. and White, G. (2012).
\newblock Self-exciting hurdle models for terrorist activity.
\newblock {\em The Annals of Applied Statistics}, 6(1):106--124.

\bibitem[Reinhart, 2018]{reinhart2018review}
Reinhart, A. (2018).
\newblock A review of self-exciting spatio-temporal point processes and their
  applications.
\newblock {\em Statistical Science}, 33(3):299--318.

\bibitem[Reinhart and Greenhouse, 2018]{reinhart2018self}
Reinhart, A. and Greenhouse, J. (2018).
\newblock Self-exciting point processes with spatial covariates: modelling the
  dynamics of crime.
\newblock {\em Journal of the Royal Statistical Society Series C: Applied
  Statistics}, 67(5):1305--1329.

\bibitem[Robbins and Monro, 1951]{robbins1951stochastic}
Robbins, H. and Monro, S. (1951).
\newblock A stochastic approximation method.
\newblock {\em The annals of mathematical statistics}, pages 400--407.

\bibitem[Rossmo, 1999]{rossmo1999geographic}
Rossmo, D.~K. (1999).
\newblock {\em Geographic profiling}.
\newblock CRC press.

\bibitem[Rumelhart et~al., 1986]{rumelhart1986learning}
Rumelhart, D.~E., Hinton, G.~E., and Williams, R.~J. (1986).
\newblock Learning representations by back-propagating errors.
\newblock {\em nature}, 323(6088):533--536.

\bibitem[Russo et~al., 2013]{russo2013criminal}
Russo, S., Roccato, M., and Vieno, A. (2013).
\newblock Criminal victimization and crime risk perception: A multilevel
  longitudinal study.
\newblock {\em Social Indicators Research}, 112:535--548.

\bibitem[Sanna~Passino et~al., 2024]{sanna2024graph}
Sanna~Passino, F., Che, Y., and Cardoso Correia~Perello, C. (2024).
\newblock Graph-based mutually exciting point processes for modelling event
  times in docked bike-sharing systems.
\newblock {\em Stat}, 13(1):e660.

\bibitem[Shchur et~al., 2021]{shchur2021neural}
Shchur, O., T{\"u}rkmen, A.~C., Januschowski, T., and G{\"u}nnemann, S. (2021).
\newblock Neural temporal point processes: A review.
\newblock {\em arXiv preprint arXiv:2104.03528}.

\bibitem[Short et~al., 2008]{short2008statistical}
Short, M.~B., D'orsogna, M.~R., Pasour, V.~B., Tita, G.~E., Brantingham, P.~J.,
  Bertozzi, A.~L., and Chayes, L.~B. (2008).
\newblock A statistical model of criminal behavior.
\newblock {\em Mathematical Models and Methods in Applied Sciences},
  18(supp01):1249--1267.

\bibitem[Stucky and Ottensmann, 2009]{stucky2009land}
Stucky, T.~D. and Ottensmann, J.~R. (2009).
\newblock Land use and violent crime.
\newblock {\em Criminology}, 47(4):1223--1264.

\bibitem[Tarzia et~al., 2018]{tarzia2018exploring}
Tarzia, L., Thuraisingam, S., Novy, K., Valpied, J., Quake, R., and Hegarty, K.
  (2018).
\newblock Exploring the relationships between sexual violence, mental health
  and perpetrator identity: a cross-sectional australian primary care study.
\newblock {\em BMC public health}, 18:1--9.

\bibitem[Veen and Schoenberg, 2008]{veen2008estimation}
Veen, A. and Schoenberg, F.~P. (2008).
\newblock Estimation of space--time branching process models in seismology
  using an em--type algorithm.
\newblock {\em Journal of the American Statistical Association},
  103(482):614--624.

\bibitem[Veličković et~al., 2018]{velickovic2018graph}
Veličković, P., Cucurull, G., Casanova, A., Romero, A., Liò, P., and Bengio,
  Y. (2018).
\newblock Graph attention networks.
\newblock In {\em International Conference on Learning Representations}.

\bibitem[Wang and Brown, 2012]{wang2012spatio}
Wang, X. and Brown, D.~E. (2012).
\newblock The spatio-temporal modeling for criminal incidents.
\newblock {\em Security Informatics}, 1:1--17.

\bibitem[Wei et~al., 2020]{wei2020distance}
Wei, N., Walteros, J.~L., and Batta, R. (2020).
\newblock On the distance between random events on a network.
\newblock {\em Networks}, 75(2):203--231.

\bibitem[Weisburd et~al., 2012]{weisburd2012criminology}
Weisburd, D., Groff, E.~R., and Yang, S.-M. (2012).
\newblock {\em The criminology of place: Street segments and our understanding
  of the crime problem}.
\newblock Oxford University Press.

\bibitem[Wu et~al., 2020]{wu2020modeling}
Wu, W., Liu, H., Zhang, X., Liu, Y., and Zha, H. (2020).
\newblock Modeling event propagation via graph biased temporal point process.
\newblock {\em IEEE Transactions on Neural Networks and Learning Systems}.

\bibitem[Wuschke and Kinney, 2018]{10.1093/oxfordhb/9780190279707.013.14}
Wuschke, K. and Kinney, J.~B. (2018).
\newblock {475Built Environment, Land Use, and Crime}.
\newblock In {\em {The Oxford Handbook of Environmental Criminology}}. Oxford
  University Press.

\bibitem[Xia et~al., 2022]{xia2022graph}
Xia, W., Li, Y., and Li, S. (2022).
\newblock Graph neural point process for temporal interaction prediction.
\newblock {\em IEEE Transactions on Knowledge and Data Engineering}.

\bibitem[Xu et~al., 2023]{xu2023semiparametric}
Xu, G., Liang, C., Waagepetersen, R., and Guan, Y. (2023).
\newblock Semiparametric goodness-of-fit test for clustered point processes
  with a shape-constrained pair correlation function.
\newblock {\em Journal of the American Statistical Association},
  118(543):2072--2087.

\bibitem[Xu and Griffiths, 2017]{xu2017shooting}
Xu, J. and Griffiths, E. (2017).
\newblock Shooting on the street: measuring the spatial influence of physical
  features on gun violence in a bounded street network.
\newblock {\em Journal of quantitative criminology}, 33:237--253.

\bibitem[Yuan et~al., 2019]{yuan2019multivariate}
Yuan, B., Li, H., Bertozzi, A.~L., Brantingham, P.~J., and Porter, M.~A.
  (2019).
\newblock Multivariate spatiotemporal hawkes processes and network
  reconstruction.
\newblock {\em SIAM Journal on Mathematics of Data Science}, 1(2):356--382.

\bibitem[Yuan et~al., 2014]{yuan2014discovering}
Yuan, N.~J., Zheng, Y., Xie, X., Wang, Y., Zheng, K., and Xiong, H. (2014).
\newblock Discovering urban functional zones using latent activity
  trajectories.
\newblock {\em IEEE Transactions on Knowledge and Data Engineering},
  27(3):712--725.

\bibitem[Zhang et~al., 2020]{zhang2020self}
Zhang, Q., Lipani, A., Kirnap, O., and Yilmaz, E. (2020).
\newblock Self-attentive hawkes process.
\newblock In {\em International Conference on Machine Learning}, pages
  11183--11193. PMLR.

\bibitem[Zhu et~al., 2021a]{zhu2021imitation}
Zhu, S., Li, S., Peng, Z., and Xie, Y. (2021a).
\newblock Imitation learning of neural spatio-temporal point processes.
\newblock {\em IEEE Transactions on Knowledge and Data Engineering},
  34(11):5391--5402.

\bibitem[Zhu et~al., 2021b]{zhu2021neural}
Zhu, S., Wang, H., Dong, Z., Cheng, X., and Xie, Y. (2021b).
\newblock Neural spectral marked point processes.
\newblock In {\em International Conference on Learning Representations}.

\bibitem[Zhu and Xie, 2022]{zhu2022spatiotemporal}
Zhu, S. and Xie, Y. (2022).
\newblock Spatiotemporal-textual point processes for crime linkage detection.
\newblock {\em The Annals of Applied Statistics}, 16(2):1151--1170.

\bibitem[Zhu et~al., 2021c]{zhu2021deep}
Zhu, S., Zhang, M., Ding, R., and Xie, Y. (2021c).
\newblock Deep fourier kernel for self-attentive point processes.
\newblock In {\em International Conference on Artificial Intelligence and
  Statistics}, pages 856--864. PMLR.

\bibitem[Zhuang and Mateu, 2019]{zhuang2019semiparametric}
Zhuang, J. and Mateu, J. (2019).
\newblock A semiparametric spatiotemporal hawkes-type point process model with
  periodic background for crime data.
\newblock {\em Journal of the Royal Statistical Society: Series A (Statistics
  in Society)}, 182(3):919--942.

\bibitem[Zipkin et~al., 2014]{zipkin2014cops}
Zipkin, J.~R., Short, M.~B., and Bertozzi, A.~L. (2014).
\newblock Cops on the dots in a mathematical model of urban crime and police
  response.
\newblock {\em Discrete and Continuous Dynamical Systems-B}, 19(5):1479--1506.

\bibitem[Zuo et~al., 2020]{zuo2020transformer}
Zuo, S., Jiang, H., Li, Z., Zhao, T., and Zha, H. (2020).
\newblock Transformer hawkes process.
\newblock In {\em International Conference on Machine Learning}, pages
  11692--11702. PMLR.

\end{thebibliography}

%% if your bibliography is in bibtex format, uncomment commands:
%\bibliographystyle{imsart-nameyear} % Style BST file
%\bibliography{bibliography}       % Bibliography file (usually '*.bib')

%% or include bibliography directly:
% \begin{thebibliography}{}
% \bibitem[\protect\citeauthoryear{???}{???}]{b1}
% \end{thebibliography}

\newpage
\appendix

\setcounter{figure}{0} \renewcommand{\thefigure}{\thesection\arabic{figure}}
\setcounter{table}{0} \renewcommand{\thetable}{\thesection\arabic{table}}
\setcounter{equation}{0} \renewcommand{\theequation}{\thesection\arabic{equation}}
% \vspace{-0.05in}

\section{Derivation of model log-likelihood}
\label{app:derivation-calculation-llk}

The log-likelihood of observing a total number of $n$ events within $[0, T] \times \mathcal{S}$ can be derived in two steps: (1) For any $1 \leq i \leq n$, compute the conditional probability density function of the $(i+1)$-th event given the previous $i$ events; (2) Use probability chain rule to get the final likelihood by multiplying $n$ conditional probability densities together. Without loss of generality, we showcase below the derivation of the $(i+1)$-th conditional probability density. For any $t \in (t_i, t_{i+1}]$, we let $F_{cl}(t) = \mathbb{P}(t_{i+1} < t, c_{i+1} = c, l_{i+1} = l | \mathcal{H}_{t_i} \cup \{(t_i, s_i, c_i \times l_i)\})$ be the cumulative probability function for the next event happened before time $t$ with mark $c\times l$. We also denote $f_{cl}(t, s) \triangleq f_{cl}(t, s|\mathcal{H}_{t_i} \cup \{(t_i, s_i, c_i \times l_i)\})$ to be the corresponding conditional probability density function for the next event with mark $c\times l$ at time $t$ and location $s$, \textit{i.e.}, $F_{cl}(t) = \int_{t_i}^{t}\int_{\mathcal{S}}f_{cl}(t, s)dsdt$. By summing over all the marks, we can define $F(t) \triangleq \sum_{c\in \mathscr{C}, l\in \mathscr{L}}F_{cl}(t) = \mathbb{P}(t_{i+1} < t | \mathcal{H}_{t_i} \cup \{(t_i, s_i, c_i \times l_i)\})$, $f(t, s) \triangleq \sum_{c\in \mathscr{C}, l\in \mathscr{L}}f_{cl}(t, s)$, and $\lambda(t, s) \triangleq \sum_{c\in \mathscr{C}, l\in \mathscr{L}}\lambda_{cl}(t, s)$. Then, if we denote $\Omega = [t, t + dt) \times B(s, \Delta s)$ to be a small neighborhood around $(t, s)$, the conditional intensity $\lambda(t, s)$ can be expressed as
\begin{equation}
    \begin{aligned}
        \lambda(t, s)|B(s, \Delta s)|dt &= \mathbb{P}\{ (t_{i+1}, s_{i+1}) \in \Omega | \mathcal{H}_{t}\} \\ &= \mathbb{P}\{(t_{i+1}, s_{i+1}) \in \Omega | \mathcal{H}_{t_i} \cup \{(t_i, s_i, c_i \times l_i)\} \cup \{t_{i+1} \geq t\}\} \\&= \frac{\mathbb{P}\{(t_{i+1}, s_{i+1}) \in \Omega, t_{n+1} \geq t | \mathcal{H}_{t_i} \cup \{(t_i, s_i, c_i \times l_i)\}\}}{\mathbb{P}\{t_{i+1} \geq t | \mathcal{H}_{t_i} \cup \{(t_i, s_i, c_i \times l_i)\}\}} \\&= \frac{f(t, s)|B(s, \Delta s)|dt}{1 - F(t)}
    \end{aligned}
    \label{eq:intensity-from-density}
\end{equation}
Integrating over $s$ we can have
\[
    \begin{aligned}
        dt \cdot \int_{\mathcal{S}}\lambda(t, s)ds = \frac{dt \cdot \int_{\mathcal{S}}f(t, s)ds}{1 - F(t)} = \frac{dF(t)}{1 - F(t)} = -d\log{(1 - F(t))}.
    \end{aligned}
\]
Replacing $t$ with $\tau$ and integrating $\tau$ over $(t_i, t)$ leads to $F(t) = 1 - \exp (-\int_{t_{i}}^{t}\int_{\mathcal{S}} \lambda(\tau, u)dud\tau)$ because $F(t_{i}) = 0$. Then we have
\begin{equation}
    f(t, s) = \lambda(t, s) \cdot \exp \left( -\int_{t_{n}}^{t}\int_{\mathcal{S}} \lambda(\tau, u)dud\tau \right)\ .
    \nonumber
\end{equation}
Since $f_{cl}(t, s)$ is proportional to $\lambda_{cl}(t, s)$, we have
\[
    \begin{aligned}
        f_{cl}(t, s) &= f(t, s) \cdot \frac{\lambda_{cl}(t, s)}{\lambda(t, s)} \\&= \lambda_{cl}(t, s) \cdot \exp \left( -\int_{t_{n}}^{t}\int_{\mathcal{S}} \lambda(\tau, u)dud\tau \right)\ .
    \end{aligned}
\]
The log-likelihood for observing the entire event sequence can be computed via the chain rule as
\[
    \begin{aligned}
        L(\theta) &= \log L(\{(t_i, s_i, c_i\times l_i)\}_{i=1}^{n}) = \log \left(\prod_{i=1}^{n} f_{c_i l_i}(t_i, s_i)\right) \\&= \sum_{i=1}^n \log \lambda_{c_il_i}\left(t_i, s_i\right)-\sum_{c\in\mathscr{C}, l\in \mathscr{L}}\int_0^T \int_{\mathcal{S}} \lambda_{cl}(t, s) ds dt,
    \end{aligned}
\]
which leads to the results in \eqref{eq:pp-log-likelihood}.

\section{Experiment details and additional results}
\label{app:additional-experiments}

\subsection{Non-parametric base event intensity estimation}
\label{app:non-para-base-estimation}

Our parameter estimation framework follows the MLE principle, as commonly adopted in point process modeling. Specifically, the MLE is applied to learn the parameters of the influence kernel $k$, while the base intensity $\mu$ is estimated in a non-parametric manner from the observed data based on kernel density estimation (KDE) \citep{reinhart2018review}. Our estimation of $\mu_{cl}$ corresponds to a piecewise-constant kernel density estimate (KDE) computed over non-overlapping spatial zones defined by event mark and urban functionality. This can be viewed as using a uniform 2D kernel with fixed support, yielding a computationally efficient approximation of the base intensity. This approach aligns with the classic MLE for point process model estimation.

\begin{table}[!h]
  \caption{Error between the estimations of $\{\mu_{cl}\}_{c\in \mathscr{C}, l\in \mathscr{L}}$ from Monte Carlo stochastic declustering and our methods.}
  % \vspace{-0.1in}
  \centering
  \resizebox{1.\linewidth}{!}{
  \begin{threeparttable}
  \begin{tabular}{ccccccc}
    \toprule
    \toprule
    % & \multicolumn{3}{c}{Traffic congestion (5 nodes)}\\
    % \cmidrule(lr){2-4} \cmidrule(lr){5-7}  \cmidrule(lr){8-10} \cmidrule(lr) {11-13}
     & MAE (rare) & MAE (frequent) & MAE (total) & MAPE (rare) & MAPE (frequent) & MAPE (total) \\
    \midrule
    % \texttt{Persistence Forecast} & 1.006 & 5.803 & 28.807 & / \\
    % \texttt{VAR} & & & & / \\
    % \midrule
    Run 1 & 0.0084 & 0.0031 & 0.0052 & 7.90\% & 5.66\% & 5.98\% \\
    Run 2 & 0.0077 & 0.0032 & 0.0049 & 7.01\% & 5.89\% & 6.05\% \\
    Run 3 & 0.0072 & 0.0029 & 0.0045 & 6.44\% & 5.13\% & 5.32\% \\
    \bottomrule
    \bottomrule
  \end{tabular}
  \begin{tablenotes}
  {\small \item *MAE and MAPE refer to the mean absolute error and mean absolute percent error between the Monte Carlo estimation and our estimation.}
  \end{tablenotes}
  \end{threeparttable}
  }
  % \vspace{-0.2in}
  \label{tab:stochastic-declustering-results}
\end{table}

To further validate our approach, we compare it with the classic non-parametric stochastic declustering approach for base intensity estimation \citep{reinhart2018review}. This method provides an accurate and spatially heterogeneous estimation of the base intensity by iteratively separating base events and those triggered by other events and using only the former for base intensity estimation via KDE. We adopt the Monte Carlo-based declustering procedure \citep{mohler2011self} for computational efficiency. Using 2015 data (7,691 events), the estimations from the two approaches closely align, as shown in Table~\ref{tab:stochastic-declustering-results}. However, our method is significantly more efficient. Our estimation is near-instantaneous compared to over 250 minutes required for more than 30 iterations of declustering. This substantial difference highlights the key advantage of our approach in its scalability, particularly in modern point process applications involving large datasets and complex models, where traditional stochastic declustering becomes computationally intractable.

\subsection{Choice of the number of training subsequences}

\begin{figure}[!t]
%\vspace{-.15in}
\centering
\begin{subfigure}[h]{.98\linewidth}
\includegraphics[width=\linewidth]{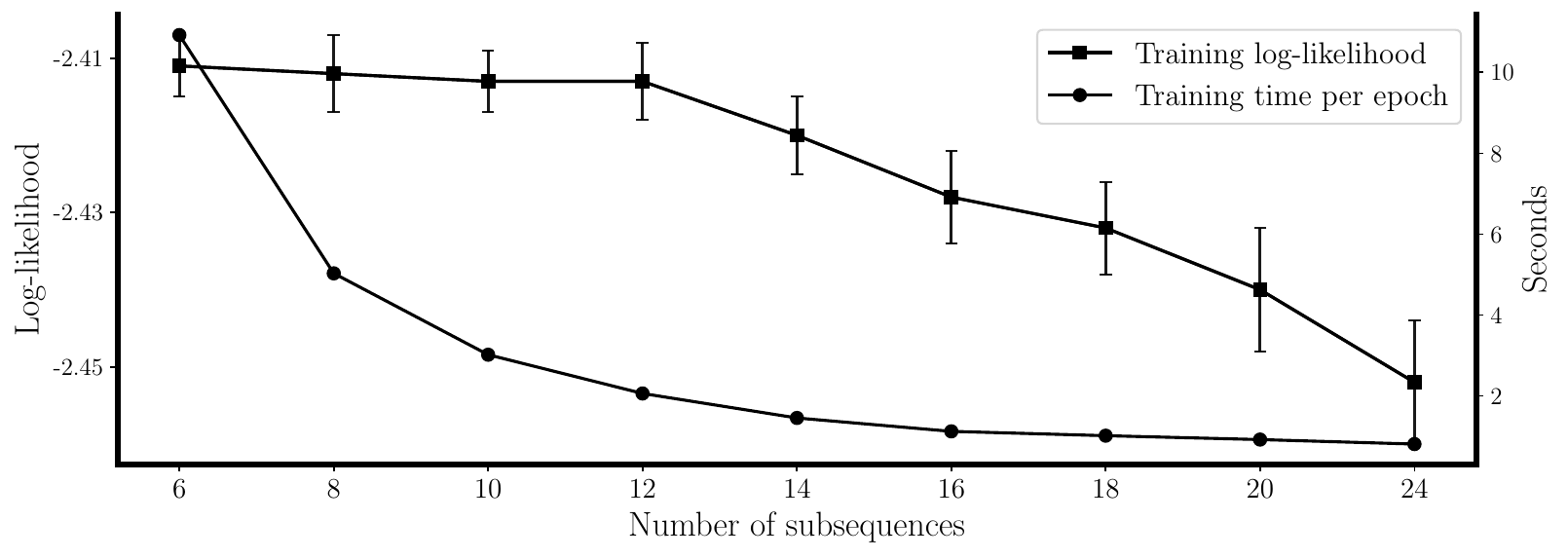}
\end{subfigure}
\caption{Model goodness-of-fit to the training data (the red line) and the computational efficiency (the blue line) with different numbers of training subsequences. The error bars on the red line indicate the standard deviation of the model training log-likelihood from three independent runs.
}
\label{fig:tradeoff-num-subseq}
% \vspace{-0.1in}
\end{figure}

The sequence splitting strategy works in our setting because, as our model assumes stationarity (empirically validated in Section~\ref{sec:model-validation}), the estimation of the influence kernel depends only on the spatio-temporal differences between pairs of events without being affected by the absolute positioning of sub-windows in time. Nonetheless, splitting the sequence reduces the total number of event pairs used in training, as it may arbitrarily split more dependent events into separate subsequences. A large number of subsequences $J$ may degrade model performance due to an insufficient number of training data. To further justify our choice of $J$, we assess the efficiency and performance of the training procedure with different numbers of event subsequences $J$ in the training set. We choose $J$ from the value set $\{6, 8, 10, 12, 14, 16, 18, 20, 24\}$. The training time per epoch (\textit{i.e.}, computation time for the entire training set) and the fitted model's log-likelihood on the training set are reported for each $J$ in Figure~\ref{fig:tradeoff-num-subseq}.

% - From the lines in the figure, we can see: entire sequence split into fewer number of subsequences, meaning a longer time window for each subsequence, can allow for a good fit to the data by the model - because the event dependencies are well preserved - but suffer from low computation efficiency - on the other hand, as we cut the entire sequence into more pieces, the computation efficiency can be improved - however, too many subsequences may break the dependency structure between events and split those events into different subsequences - thus, the model will underfit the data - We choose the number of subsequences to be $12$ to strike the balance between model's goodness-of-fit and the training efficiency.

As observed, partitioning the entire sequence into fewer subsequences (\textit{i.e.}, longer time window for each subsequence) allows the model to better fit the data. Note that a longer time window for subsequences means more preservation of the dependencies among events. This preservation is crucial for achieving a good fit, as it ensures that dependent events are analyzed within the same context. However, a longer length of each subsequence demands higher computational complexity for the log-likelihood function in \eqref{eq:pp-log-likelihood}, thus reducing the model training efficiency.
On the other hand, a sufficiently large number of subsequences reduces the complexity of evaluating the log-likelihood and enhances computational efficiency, while resulting in an underfitting of the data, failing to capture essential patterns of dependencies among crime events.

Our choice of $J=12$ justified by the performance metrics in Figure~\ref{fig:tradeoff-num-subseq} leads to a $120$-day window length, which is more than three times larger than the learned temporal decay scale (approximately 30.77 days). This ensures the preservation of the vast majority of event pairs with non-zero dependencies within individual subsequences, striking a balance between the computational efficiency and the model's goodness-of-fit to the data. 
We also note that while this splitting strategy is appropriate and unbiased under stationary assumptions, its applicability may be more limited in non-stationary settings where model parameters or dynamics vary over time or space. In such cases, the loss of cross-sequence dependencies could affect the estimation reliability, and additional justification through theoretical or empirical validation would be required.

\subsection{Baseline descriptions}

We compare our model with the following four baselines: 

\begin{itemize}
    \item The persistence forecast (\texttt{Persistent}) is a simple and straightforward forecasting technique where the future value is predicted to be the same as the most recent observed value. In our experiments, the number of events with a specified mark in the next week $t+1$ is predicted as the number of events with the same mark observed in the current week $t$.

    \item The Vector Autoregression (\texttt{VAR}) is a statistical model used to capture the linear dependencies among multiple time series. VAR generalizes the univariate autoregressive model (AR) by modeling each variable in the system as a linear combination of past values of itself and past values of all the other variables in the system. Specifically, denoting the variable vector as $y \in \mathbb{R}^d$ and its value at time $t$ as $y_t$, the linear relationship between future values and past values is expressed as $$y_t = C + \sum_{i=1}^{p}A_iy_{t-i} + \epsilon_t.$$ Here $C \in \mathbb{R}^d$ is a constant vector, $A_i$ are coefficient matrices, and $\epsilon_t$ is a white noise vector.

    \item The Epidemic-type aftershock sequence (\texttt{ETAS}) model is a benchmark point process model for modeling spatio-temporal discrete event data. The original ETAS only models the time and location of the event without considering the event type. Here, we slightly modify the original model by incorporating a set of coefficients to account for the interactions between different event marks. Specifically, the influence kernel takes the form of a diffusion-type kernel as
    \[
    \begin{aligned}
        k(t', t, s', s, c'\times l', c\times l) = \frac{\eta_{cl, c'l'} e^{-\beta(t - t^\prime)}}{2\pi \sqrt{|\Sigma|}(t - t^\prime)} \cdot \exp{\left \{ -\frac{(s - s^\prime)^\top \Sigma^{-1} (s - s^\prime)}{2(t - t ^\prime)} \right \}}.
    \end{aligned}
    \]
    Here $\mu \geq 0$ is the base event intensity, $\Sigma = \mathrm{diag}(\sigma_x^2, \sigma_y^2)$ is a two-dimensional diagonal matrix representing the covariance of the spatial correlation, $\beta > 0$ is the decaying rate, and $\eta_{cl, c'l'} > 0$ controls the magnitude of the influence from past events. We use the same estimation strategy for the base intensity $\mu$ and estimate other parameters $\{\sigma_x, \sigma_y, \beta, \eta_{cl, c'l'}\}$ using the same SGD with regard to model likelihood.

    \item The \texttt{STNPP} without GAT (\texttt{STNPP-GAT}) is an ablated variant of our model where we remove the GAT architecture and directly estimate the coefficients $\{\alpha_{cl, c'l'}\}_{c,c'\in\mathscr{C}, l,l'\in \mathscr{L}}$ using SGD. The goal of comparing our model to \texttt{STNPP-GAT} is to showcase how the integration of the GAT architecture enhances our ability to discern the intricate patterns of mark interactions. This improvement facilitates the identification of closely related marks and yields more precise predictions.
\end{itemize}

\subsection{Next-event prediction}

\begin{table}[!t]
  \caption{Model performance on next event prediction: time, location, and type.}
  \centering
  \resizebox{.8\linewidth}{!}{
  \begin{threeparttable}
  \begin{tabular}{ccccc}
    \toprule
    \toprule
    % & \multicolumn{3}{c}{Traffic congestion (5 nodes)}\\
    % \cmidrule(lr){2-4} \cmidrule(lr){5-7}  \cmidrule(lr){8-10} \cmidrule(lr) {11-13}
    Model & Time MAE ($\downarrow$) & Location MAE ($\downarrow$) & Type Accuracy ($\uparrow$) \\
    \midrule
    \texttt{Persistent} & 0.036 & 3.754 & 0.200 \\
    \texttt{ETAS} & 0.038 & 3.649 & 0.210 \\
    \texttt{STNPP-GAT} & 0.032 & 3.572 & 0.283 \\
    \texttt{STNPP} & {\bf 0.027} & {\bf 3.231} & {\bf 0.302} \\
    \bottomrule
    \bottomrule
  \end{tabular}
  % \begin{tablenotes}
  % \item *Numbers in parentheses are standard errors for three independent runs.
  % \end{tablenotes}
  \end{threeparttable}
  }
  % \vspace{-0.2in}
  \label{tab:next-event-prediction-results}
\end{table}

One of the important criteria for assessing the real-world applicability of a point process model is the model's predictive accuracy on next-event forecasting. To this end, we conduct an experiment focused on next-event prediction. Specifically, we sample 1,000 event sequences from the 2019 data as the testing set, each starting at the first event in 2019 and ending at a randomly selected event that occurs after July 1st, 2019. For each sequence, we treat the final event as the prediction target and use its preceding history as input to the fitted models. We evaluate our model \texttt{STNPP} and other baselines (\texttt{Persistent}, \texttt{ETAS}, \texttt{STNPP-GAT}) on this prediction task to predict the time, type, and location of the last event in each sequence. In particular, the persistent prediction method refers to the naive prediction by copying the inter-arrival time, type, and location of the most recent past event to the predicted event. The time series model \texttt{VAR} we compared in the original paper does not apply to the individual-event-level prediction.
    
As shown in Table~\ref{tab:next-event-prediction-results}, our proposed model \texttt{STNPP} achieves the lowest MAE in predicting both time and spatial location of the event, and the highest accuracy for predicting the event type. These results highlight our model's strength in capturing the temporal dynamics, spatial dependencies over the street network, and structured interactions between crime-landmark marks. Notably, the improvement over \texttt{STNPP-GAT} suggests that incorporating GNN-based mark interaction modeling also leads to an enhancement in model short-term forecasting.

\subsection{Parameter identifiability}

The interactions between different event marks are captured by the coefficients $\alpha_{cl,c'l'}$, modeled as the product of the strength and chance variable $\alpha_{cl,c'l'} = a_{cl,c'l'} p_{cl,c'l'}$.
To ensure the parameter identifiability and model estimation convergence, we impose constraints on the strength and chance variable. First, we require the $a_{cl,c'l'}$ and $p_{cl,c'l'}$ to be non-negative, and the $p_{cl,c'l'}$ to be in the $[0, 1]$ interval as a probability. Second, the $\{p_{cl,c'l'}\}$ are constrained to form a valid probability distribution across all potential triggering marks $c' \times l'$ for a given mark $c \times l$, \textit{i.e.}, $\sum_{c'\in \mathscr{C}, l'\in \mathscr{L}}p_{cl,c'l'} = 1$. This is ensured via a softmax normalization \eqref{eq:normalized-attention-weights} in the design of GAT for modeling the chance variable. These constraints help disambiguate the scaling between $a_{cl,c'l'}$ and $p_{cl,c'l'}$.

\begin{figure}[!t]
\centering
\begin{subfigure}[h]{.98\linewidth}
\includegraphics[width=\linewidth]{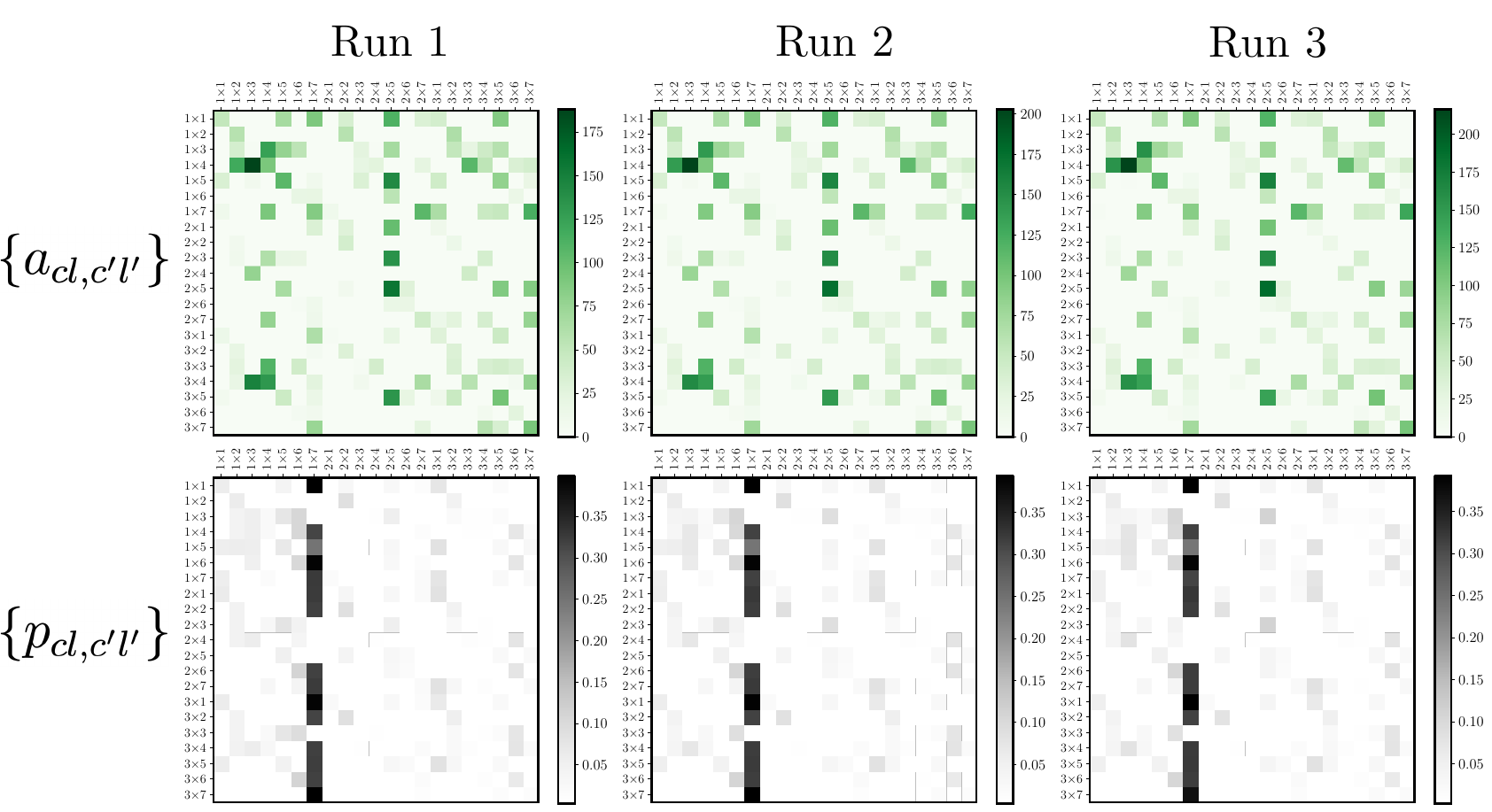}
\end{subfigure}
\caption{Learned $\{a_{cl,c'l'}\}$ and $\{p_{cl,c'l'}\}$ under different random initializations.
}
\label{fig:a-p_matrices}
\end{figure}
    
Admittedly, training neural networks involves solving a highly non-convex optimization problem (specifically, maximizing the likelihood \eqref{eq:pp-log-likelihood}), which may not yield a unique global solution due to the inherent non-convexity. Nevertheless, as observed in many deep learning contexts, sufficiently large neural networks possess strong expressive power in representing $p_{cl,c'l'}$, and the solutions for $p_{cl,c'l'}$ and $a_{cl,c'l'}$ appear robust, as we observed in our empirical results. We trained our model three times using the same architecture and training data (2015–2018 crime data, as described in the main paper), but with different model initializations in each run by setting different random seeds. As shown in Figure~\ref{fig:a-p_matrices}, the learned $\{a_{cl,c'l'}\}$ and $\{p_{cl,c'l'}\}$ have similar values and structures across three independent runs. This suggests that the estimated parameters are not sensitive to initialization and remain consistent across different numerical solvers and initialization schemes.

\end{document}